\documentclass[aps,prd,preprintnumbers,groupedaddress,nofootinbib,amssymb,notitlepage,eqsecnum]{revtex4-1}
\usepackage{here}
\usepackage[dvipdfmx]{graphicx}
\usepackage{amsmath,amsthm,amssymb}
\usepackage{bm}
\usepackage{color}

\usepackage{amsfonts}
\usepackage{dcolumn}
\usepackage{hyperref}
\allowdisplaybreaks[1]
\usepackage{stackengine}

\newcommand{\be}{\begin{equation}}  
\newcommand{\ee}{\end{equation}}
\newcommand{\ba}{\begin{eqnarray}}
\newcommand{\ea}{\end{eqnarray}}

\newcommand{\rd}{{\rm d}}
\newcommand{\hr}{\hat{r}}

\newcommand{\bem}{\begin{bmatrix}}
\newcommand{\eem}{\end{bmatrix}}
\newcommand{\Mpl}{M_{\rm Pl}}

\allowdisplaybreaks

\begin{document}

\preprint{WUCG-23-02}
\title{Black hole perturbations in Maxwell-Horndeski theories}

\author{Ryotaro Kase$^{1}$ and Shinji Tsujikawa$^{2}$}

\affiliation{
$^1$Department of Physics, Faculty of Science, 
Tokyo University of Science, 1-3, Kagurazaka,
Shinjuku-ku, Tokyo 162-8601, Japan\\
$^2$Department of Physics, Waseda University, 3-4-1 Okubo, 
Shinjuku, Tokyo 169-8555, Japan}

\begin{abstract}

We study the linear stability of black holes in Maxwell-Horndeski 
theories where a $U(1)$ 
gauge-invariant vector field is coupled to a scalar field with the Lagrangian of full Horndeski theories. 
The perturbations on a static and spherically symmetric background 
can be decomposed into odd- and even-parity modes under the expansion of 
spherical harmonics with 
multipoles $l$. 
For $l \geq 2$, the odd-parity sector contains two propagating degrees of freedom associated with the gravitational and vector field perturbations. 
In the even-parity sector, there are three dynamical perturbations 
arising from the scalar field besides the gravitational and vector field perturbations. For these five propagating degrees of freedom, 
we derive conditions for the absence of ghost/Laplacian 
stabilities along the radial and angular directions. 
We also discuss the stability of black holes for $l=0$ and $l=1$, in which case no additional conditions 
are imposed to those obtained 
for $l \geq 2$. 
We apply our general results to Einstein-Maxwell-dilaton-Gauss-Bonnet 
theory and Einstein-Born-Infeld-dilaton gravity and show that hairy black hole 
solutions present in these theories can be consistent with all the 
linear stability conditions. 
In regularized four-dimensional Einstein-Gauss-Bonnet gravity 
with a Maxwell field, however, exact charged black hole solutions known in 
the literature are prone to 
instabilities of even-parity perturbations besides a strong coupling problem with a vanishing kinetic term of the radion mode.

\end{abstract}

\date{\today}


\maketitle

\section{Introduction}
\label{introsec}

Black holes (BHs) are the fundamental object arising as a solution to 
the Einstein field equation in General Relativity (GR). 
On a static and spherically symmetric background, the 
Schwarzschild geometry characterized by a single mass parameter $M$ 
is a unique asymptotically flat solution in GR without matter. 
In the presence of a Maxwell field, there is a static 
BH with an electric charge $q$ 
known as a Reissner-Nordstr\"{o}m (RN) solution. 
Allowing the rotation of BHs leads to a Kerr solution containing an 
angular momentum $J$.
In Einstein-Maxwell theory without additional matter, there is a uniqueness 
theorem stating that stationary and asymptotically-flat BHs are characterized only by three parameters, i.e., 
$M$, $q$, and $J$ \cite{Israel:1967wq,Carter:1971zc,Ruffini:1971bza,Hawking:1971vc}.

If we take an extra degree of freedom into account, it is possible 
to have additional BH ``hairs'' to those present in GR without matter. 
For a minimally coupled canonical scalar field 
$\phi$ \cite{Hawking:1971vc,Bekenstein:1972ny}
and k-essence \cite{Graham:2014mda} as well as for 
a nonminimally coupled scalar 
field with the Ricci scalar $R$ of the form
$G_4(\phi)R$ \cite{Hawking:1972qk,Bekenstein:1995un,Sotiriou:2011dz,Faraoni:2017ock}, 
it is known that BHs do not have scalar hairs. 
If the scalar field is coupled to 
a Gauss-Bonnet (GB) term $R_{\rm GB}^2$
of the form $\xi(\phi) R_{\rm GB}^2$ \cite{Zwiebach:1985uq,Antoniadis:1993jc,Gasperini:1996fu}, 
where $\xi(\phi)$ is a function of $\phi$, the existence of 
asymptotically flat hairy BH solutions was shown for the dilatonic 
coupling $\xi(\phi) \propto e^{-\lambda \phi}$~\cite{Kanti:1995vq,Alexeev:1996vs,Kanti:1997br,Chen:2006ge,Guo:2008hf,Guo:2008eq,Pani:2009wy,Ayzenberg:2014aka,Maselli:2015tta,Kleihaus:2011tg,Kleihaus:2015aje} and the linear coupling~$\xi(\phi) \propto \phi$~\cite{Sotiriou:2013qea,Sotiriou:2014pfa}. 
It was also found that, for the scalar-GB coupling $\xi(\phi)$ 
with even power-law functions of $\phi$, 
a phenomenon called spontaneous scalarization of BHs can 
occur \cite{Doneva:2017bvd,Silva:2017uqg,Antoniou:2017acq,Blazquez-Salcedo:2018jnn,Minamitsuji:2018xde,Silva:2018qhn,Macedo:2019sem,Doneva:2021tvn}, analogous to spontaneous scalarization of neutron stars 
induced by a nonminimal coupling with the Ricci scalar \cite{Damour:1993hw}.

The scalar-GB coupling mentioned above belongs to a subclass of Horndeski 
theories with second-order Euler equations 
of motion \cite{Horndeski,KYY}.
If we consider a time-independent scalar field on the static and spherically 
symmetric background, there are some other subclasses of Horndeski 
theories in which hairy BH solutions are present. 
One example is a scalar nonminimal derivative coupling~$\phi\,G_{\mu \nu} \nabla^{\mu} \nabla^{\nu}\phi$ to the Einstein tensor $G_{\mu \nu}$, 
in which case non-asymptotically flat BH solutions are present \cite{Rinaldi:2012vy,Anabalon:2013oea,Minamitsuji:2013ura,Cisterna:2014nua,Kolyvaris:2011fk,Minamitsuji:2014hha}. 
However, it was recently recognized that these solutions are 
unstable against linear perturbations around the BH  
horizon \cite{Minamitsuji:2022mlv}.
In so-called regularized four-dimensional Einstein-Gauss-Bonnet (4DEGB) 
theory \cite{Glavan:2019inb} where the GB coupling $\hat{\alpha}_{\rm GB}R_{\rm GB}^2$ in a $D$-dimensional spacetime is rescaled as $\hat{\alpha}_{\rm GB} \to \alpha_{\rm GB}/(D-4)$ on a $(D-4)$-dimensional maximally symmetric 
flat space \cite{Lu:2020iav,Kobayashi:2020wqy}, 
there exists an exact hairy BH solution respecting the asymptotic flatness. 
The 4DEGB gravity also belongs to a subclass of Horndeski theories 
with the scalar field playing the role of a radion \cite{Fernandes:2022zrq}, so 
the linear stability conditions derived in Refs.~\cite{Kobayashi:2012kh,Kobayashi:2014wsa,Kase:2021mix} for full Horndeski theories can be applied to this case 
as well. The recent study \cite{Tsujikawa:2022lww} showed that the exact BH solution present in 4DEGB gravity is not only unstable but also plagued by 
a strong coupling problem.  

The instabilities of BHs found in nonminimal derivative coupling and 4DEGB 
theories are related to a finite 
scalar field kinetic term 
$X=-(1/2)\nabla^{\mu} \phi \nabla_{\mu} \phi$ 
on the horizon \cite{Minamitsuji:2022mlv}. 
If we try to search for asymptotically flat hairy BHs
with a static scalar field in full Horndeski theories, 
models with regular coupling functions $G_{2,3,4,5}$ 
of $\phi$ and $X$ generally result in no-hair Schwarzschild 
solutions \cite{Minamitsuji:2022vbi}. 
The exceptional case is the scalar-GB coupling 
$\xi(\phi)R_{\rm GB}^2$ mentioned above, in which case 
the corresponding hairy BHs can be consistent with 
all the linear stability conditions 
in a small GB coupling regime.
For a scalar field having the dependence of time $t$ in the form 
$\phi=q_c t+\Psi(r)$, where $q_c$ is 
a constant and $\Psi(r)$ is a function of the radial coordinate, 
it is known that a stealth Schwarzschild solution is also present \cite{Babichev:2013cya,Kobayashi:2014eva}. 
It is still fair to say that the construction of asymptotically flat hairy BHs free from instabilities is limited in the framework of Horndeski theories, 
especially for a time-independent scalar 
field.

If we consider an electromagnetic tensor $F_{\mu \nu}$ coupled to the 
scalar field $\phi$, there are more possibilities for realizing hairy BHs. 
{}From the theoretical perspective, heterotic string theory gives rise to 
a coupling between the dilaton field 
$\phi$ and Maxwell field strength 
$F=-F_{\mu \nu}F^{\mu \nu}/4$. 
In Einstein-Maxwell-dilaton theory given by the Lagrangian 
${\cal L}=R+4X+4e^{-2\phi}F$, 
Gibbons and Maeda (GM) \cite{Gibbons:1987ps} and 
Garfinkle, Horowitz, and Strominger (GHS) \cite{Garfinkle:1990qj} found 
charged hairy BH solutions 
with a nonvanishing dilaton.
The dilatonic hair appears as a result of the coupling 
with the electromagnetic field. We note that, 
for scalar-vector couplings $\xi(\phi)F$ with even power-law functions of $\phi$ in $\xi$, the RN BH can trigger tachyonic instability to evolve into a 
scalarized charged 
BH \cite{Herdeiro:2018wub,Fernandes:2019rez,Myung:2018jvi,Blazquez-Salcedo:2020nhs,Konoplya:2019goy}.
The low energy effective action in string theory also 
contains a coupling between the dilaton and the GB term  
as a next-to-leading order term of the inverse string 
tension $\alpha'$. In the presence of the dilatonic coupling 
with both Maxwell and GB terms, Mignemi and 
Stewart \cite{Mignemi:1992nt} showed the 
existence of hairy BH solutions by using 
an expansion in terms of the 
small coupling $\alpha'$ 
(see Refs.~\cite{Torii:1996yi,Alexeyev:2009kx} for related works).

In 4DEGB gravity, the hairy BH said before
corresponds to an exact solution without a Maxwell field. 
Analogous to the case of string theory, 
one can incorporate an electromagnetic field in the four-dimensional effective action.  
Indeed, there exists an exact charged BH solution in 4DEGB 
gravity \cite{Fernandes:2020rpa}, which is analogous to 
those derived in higher-dimensional 
setups \cite{Wiltshire:1985us,Wiltshire:1988uq,Cai:2001dz}.
It is not yet clarified whether this charged BH has the problems 
of instability and strong coupling mentioned above.

In open string theory, there are possible corrections to the Maxwell action arising from couplings of the Abelian gauge field to bosonic strings \cite{Tseytlin:1986ti,Fradkin:1985qd,Abouelsaood:1986gd}. 
The tree-level effective electromagnetic action coincides with 
a nonlinear action of Born and Infeld (BI) given by the Lagrangian 
${\cal L}=(4/b^2) ( 1-\sqrt{1-2b^2 F})$ \cite{Born:1934gh}. 
At leading order in the expansion with respect to a small coupling constant $b$, 
the BI action recovers the Maxwell Lagrangian ${\cal L}=4F$. 
In the four-dimensional Einstein-BI gravity, there is an exact BH 
solution whose metric differs from the RN 
solution \cite{Fernando:2003tz,Cai:2004eh,Dey:2004yt}.
One can deal with such a nonlinear electromagnetism by considering 
a general function of $G_2(F)$ in 
the Lagrangian.
If there is a scalar field $\phi$ coupled with the Maxwell field, 
the Lagrangian can be further extended to the form $G_2(\phi, X, F)$.  
Indeed, in Einstein-Born-Infeld-dilaton gravity where the dilaton is coupled to the BI field, the existence of 
hairy BH solutions is also known 
\cite{Clement:2000ue,Tamaki:2001vv,Yazadjiev:2005za,Sheykhi:2006ji,Stefanov:2007qw}.

In this paper, we study the stability of static and spherically 
symmetric BH solutions in four-dimensional 
Maxwell-Horndeski theories where the scalar field $\phi$ with
the Horndeski Lagrangian is coupled 
to a $U(1)$ gauge-invariant vector field 
through the coupling $G_2(\phi, X, F)$. 
A similar study was performed in Ref.~\cite{Gannouji:2021oqz} 
for the Lagrangian ${\cal L}=G_2(\phi, X, F)+G_4(\phi)R$, but 
our analysis is more general in that the scalar field sector is 
described by the full Horndeski Lagrangian.  
By doing this, we can accommodate the stabilities of hairy 
BH solutions present in all the theories 
mentioned above, especially those containing the GB term.
 
We decompose the types of perturbations into the odd- 
and even-parity sectors and derive all the linear stability 
conditions of five dynamical perturbations. 
In particular we will derive the propagation speeds of 
even-parity perturbations along the angular direction 
in the limit of large multipoles $l$, which are missing  
in most of the papers about BH perturbations 
in the literature \cite{Kobayashi:2014wsa,Gannouji:2021oqz}.
We note that, in full Horndeski theories with a perfect fluid, 
all the linear stability conditions including the angular propagation 
speeds were derived in Ref.~\cite{Kase:2021mix}, which can be applied to the BH case as well (see also Ref.~\cite{Kase:2020qvz}). 
Indeed, the angular Laplacian stability is important to 
exclude hairy BHs arising in nonminimal derivative coupling 
theories \cite{Minamitsuji:2022mlv,Minamitsuji:2022mlv}
and in 4DEGB gravity \cite{Tsujikawa:2022lww}. 
Neutron stars with scalar hairs present 
in the same theories are also prone to similar instability problems \cite{Kase:2020yjf,Kase:2021mix,Minamitsuji:2022tze}.

After deriving all the linear stability conditions of odd- and even-parity perturbations in Maxwell-Horndeski 
theories, we will apply them to concrete hairy BH solutions present in 
Einstein-Maxwell-dilaton theory, 
Einstein-BI-dilaton gravity, 
Einstein-Maxwell-dilaton-GB theory, 
and 4DEGB gravity. While the first three theories allow the existence of 
charged BHs consistent with all the linear stability conditions, 
the exact charged BH solution present 
in 4DEGB gravity suffers from Laplacian instability of even-parity 
perturbations as well as the strong coupling problem. 
The nature of instabilities is similar to what was found for the 
uncharged exact BH solution in 4DEGB gravity \cite{Tsujikawa:2022lww}.
Thus, our stability criteria in 
Maxwell-Horndeski theories are useful to 
exclude some BH solutions or constrain 
allowed parameter spaces.
The second-order actions of odd- and 
even-parity perturbations and resulting field equations of motion can be also applied to the computation of BH quasinormal modes. 

This paper is organized as follows. 
In Sec.~\ref{scasec}, we derive the field equations of motion in 
Maxwell-Horndeski theories on the static and spherically 
symmetric background. 
In Sec.~\ref{oddsec}, we obtain conditions for the absence of 
ghost/Laplacian instabilities in the odd-parity sector and show that 
the propagation of vector field perturbation is luminal with 
the other stability conditions similar to those in Horndeski theories. 
In Sec.~\ref{evensec}, we derive the second-order action of 
even-parity perturbations and clarify 
how the vector field perturbation 
affects the linear stability conditions. Since the number of 
dynamical degrees of freedom (DOFs) depends on the multipole $l$ 
in the expansion of spherical harmonics, we discuss the cases 
$l\geq 2$, $l=0$, and $l=1$, in turn.
In Sec.~\ref{appsec}, we apply our general results to the stability of 
hairy BHs present in several classes of theories mentioned above. 
Sec.~\ref{consec} is devoted to conclusions.

\section{Maxwell-Horndeski theories}
\label{scasec}

We consider a scalar field $\phi$ in the framework of  
Horndeski theories with second-order Euler equations 
of motion \cite{Horndeski}. 
We also incorporate a $U(1)$ gauge-invariant vector field $A_{\mu}$ with the field strength 
tensor $F_{\mu \nu}=\nabla_{\mu} A_{\nu}-\nabla_{\nu} A_{\mu}$, where 
$\nabla_{\mu}$ is a covariant-derivative operator. 
The vector field Lagrangian depends on a scalar quantity 
\be
F \equiv -\frac{1}{4} F_{\mu \nu} F^{\mu \nu}\,.
\ee
We allow the existence of couplings between the scalar and vector fields 
of the form $G_2(\phi, X, F)$, where $G_2$ is a function of $\phi$,  
$X=-(1/2)\nabla^{\mu} \phi \nabla_{\mu} \phi$, and $F$. 
The action of Maxwell-Horndeski theories is given by 
\ba
\hspace{-1.2cm}
& &
{\cal S}=
\int {\rm d}^4 x \sqrt{-g}\,
\bigg\{ G_2(\phi,X,F)-G_{3}(\phi,X)\square\phi 
+G_{4}(\phi,X)\, R +G_{4,X}(\phi,X)\left[ (\square \phi)^{2}
-(\nabla_{\mu}\nabla_{\nu} \phi)
(\nabla^{\mu}\nabla^{\nu} \phi) \right] \nonumber \\
\hspace{-1.2cm}
& &
\quad +G_{5}(\phi,X)G_{\mu \nu} \nabla^{\mu}\nabla^{\nu} \phi
-\frac{1}{6}G_{5,X}(\phi,X)
\left[ (\square \phi )^{3}-3(\square \phi)\,
(\nabla_{\mu}\nabla_{\nu} \phi)
(\nabla^{\mu}\nabla^{\nu} \phi)
+2(\nabla^{\mu}\nabla_{\alpha} \phi)
(\nabla^{\alpha}\nabla_{\beta} \phi)
(\nabla^{\beta}\nabla_{\mu} \phi) \right]\bigg\},
\label{action}
\ea
where $g$ is a determinant of the metric tensor 
$g_{\mu \nu}$, and $G_3, G_4, G_5$ are functions of 
$\phi$ and $X$.
We use the notations  
$\square \phi \equiv \nabla^{\mu}\nabla_{\mu} \phi$ and 
$G_{j,\phi} \equiv \partial G_j/\partial \phi$, 
$G_{j,X} \equiv \partial G_j/\partial X$, 
$G_{j,\phi X} \equiv \partial^2 G_j/(\partial X \partial \phi)$ ($j=2,3,4,5$), 
and so on. 
The action (\ref{action}) is invariant under the shift $A_{\mu} \to A_{\mu}+\partial_{\mu}\chi$, 
so the theory respects a $U(1)$ gauge symmetry. 
Introducing the gauge-invariant vector field $A_{\mu}$ to the Horndeski action 
gives rise to two additional dynamical  DOFs to those in Horndeski theories 
(one scalar and two tensor modes). 
Hence the total propagating DOFs are five in 
Maxwell-Horndeski theories\footnote{If we consider generalized 
Proca (GP) theories \cite{Heisenberg:2014rta,Tasinato:2014eka,BeltranJimenez:2016rff} 
with a $U(1)$-symmetry breaking gauge field, 
there is an additional longitudinal propagation of the vector field. 
It is known that there are hairy BH solutions in GP 
theories \cite{Chagoya:2016aar,Minamitsuji:2016ydr,Babichev:2017rti,Heisenberg:2017xda,Heisenberg:2017hwb}, 
but our analysis in this paper does not accommodate such cases. 
Readers may refer to Refs.~\cite{Kase:2018voo,Kase:2018owh,Baez:2022rdz,Garcia-Saenz:2021uyv} 
for BH perturbations (mostly in the odd-parity sector) in GP theories 
and its extensions.}.

In this section, we derive the background equations of motion on 
a static and spherically symmetric spacetime given by 
the line element 
\be
{\rm d}s^{2} =-f(r) {\rm d}t^{2} +h^{-1}(r) {\rm d}r^{2} + 
r^{2} \rd \Omega^2\,,
\label{metric}
\ee
where $\rd \Omega^2={\rm d}\theta^{2}
+\sin^{2}\theta\, {\rm d} \varphi^{2}$, 
and $t$, $r$ and $(\theta,\varphi)$ represent the time, radial, and angular coordinates, respectively, and $f$ and $h$ are functions of $r$. Since we are interested in the 
stability of BHs outside the horizon, we will consider positive 
values of $f$ and $h$.
On the background (\ref{metric}), we consider a time-independent scalar field 
with the radial dependence
\be
\phi=\phi(r)\,.
\label{phir}
\ee
As we mentioned in Introduction, Maxwell-Horndeski theories given 
by the action (\ref{action}) can accommodate a variety 
of hairy BH solutions known in the literature.
For the vector field, we consider the following 
configuration \cite{DeFelice:2016cri}
\be
A_{\mu}=[A_0(r), A_1(r), 0, 0]\,.
\ee
In the $U(1)$ gauge-invariant theory under consideration now, 
the longitudinal mode $A_1(r)$ does not contribute to 
the background equations.
The scalar quantities $X$ and $F$ reduce, respectively, to 
\be
X=-\frac{1}{2} h \phi'^2\,,\qquad 
F=\frac{h}{2f}A_0'^2\,,
\label{BGF}
\ee
where a prime represents the derivative with respect to $r$.

Varying the action (\ref{action}) with respect to $g_{\mu\nu}$, 
the (00), (11), (22) components of gravitational field equations 
of motion are given, respectively, by 
\ba
{\cal E}_{00}&\equiv&
\left(C_1+\frac{C_2}{r}+\frac{C_3}{r^2}\right)\phi''
+\left(\frac{\phi'}{2h}C_1+\frac{C_4}{r}+\frac{C_5}{r^2}\right)h'
+C_6+\frac{C_7}{r}+\frac{C_8}{r^2}-\frac{h}{f} G_{2,F}A_0'^2=0
\,,\label{back1}\\
{\cal E}_{11}&\equiv&
-\left(\frac{\phi'}{2h}C_1+\frac{C_4}{r}+\frac{C_5}{r^2}\right) \frac{hf'}{f}
+C_9-\frac{2\phi'}{r}C_1-\frac{1}{r^2}\left[\frac{\phi'}{2h}C_2+(h-1)C_4\right]
+\frac{h}{f} G_{2,F}A_0'^2=0
\,,\label{back2}\\
{\cal E}_{22}&\equiv&
\left[\left\{C_2+\frac{(2h-1)\phi'C_3+2hC_5}{h\phi' r}\right\}\frac{f'}{4f}+C_1+\frac{C_2}{2r}\right]\phi''
+\frac{1}{4f}\left(2hC_4-\phi'C_2+\frac{2hC_5-\phi'C_3}{r}\right)\left(f''-\frac{f'^2}{2f}\right)\notag\\
&&
+\left[C_4+\frac{2h(2h+1)C_5-\phi'C_3}{2h^2r}\right]\frac{f'h'}{4f}
+\left(\frac{C_7}{4}+\frac{C_{10}}{r}\right)\frac{f'}{f}
+\left(\frac{\phi'}{h}C_1+\frac{C_4}{r}\right)\frac{h'}{2}+C_6+\frac{C_7}{2r}
=0\,,\label{back3}
\ea
where a prime represents the derivative with respect to $r$, and
the coefficients are given by
\ba
&&
C_1=-h^2 (G_{3,X}-2 G_{4,\phi X} ) \phi'^2-2 G_{4,\phi} h\,,\notag\\
&&
C_2=2 h^3 ( 2 G_{4,XX}-G_{5,\phi X} ) \phi'^3-4 h^2 ( G_{4,X}-G_{5,\phi} ) \phi'\,,\notag\\
&&
C_3=-h^4G_{5,XX} \phi'^4+h^2G_{5,X}  ( 3 h-1 ) \phi'^2\,,\notag\\
&&
C_4=h^2 ( 2 G_{4,XX}-G_{5,\phi X} ) \phi'^4+h ( 3 G_{5,\phi}-4 G_{4,X} ) \phi'^2-2 G_4\,,\notag\\
&&
C_5=-\frac12 \left[G_{5,XX} h^3{\phi'}^{5}- hG_{5,X}  ( 5 h-1 ) \phi'^3\right]\,,\notag\\
&&
C_6=h ( G_{3,\phi}-2 G_{4,\phi\phi} ) \phi'^2+G_2\,,\notag\\
&&
C_7=-2 h^2 ( 2 G_{4,\phi X}-G_{5,\phi\phi} ) \phi'^3-4 G_{4,\phi} h\phi'\,,\notag\\
&&
C_8=G_{5,\phi X} h^3\phi'^4-h ( 2 G_{4,X} h-G_{5,\phi} h-G_{5,\phi} ) \phi'^2-2 G_4  ( h-1 )\,,\notag\\
&&
C_9=-h ( G_{2,X}-G_{3,\phi} ) \phi'^2-G_2\,,\notag\\
&&
C_{10}=\frac12 G_{5,\phi X} h^3\phi'^4-\frac12 h^2 ( 2 G_{4,X}-G_{5,\phi} ) \phi'^2-G_4 h\,.
\ea
The scalar field equation of motion following from the variation of 
(\ref{action}) with respect to $\phi$ gives 
\be
\frac{1}{r^2} \sqrt{\frac{h}{f}} \left( r^2 \sqrt{\frac{f}{h}} J^r 
\right)'+\frac{\partial {\cal E}}{\partial\phi}=0\,,
\label{Ephi}
\ee
where
\ba
J^r&=&\left(C_1+\frac{C_2}{r}+\frac{C_3}{r^2}\right)\frac{f'}{2f}
-\frac{C_6+C_9}{\phi'}+\frac{2}{r}C_1+\frac{1}{r^2}\left(\frac{1+h}{2h}C_2
-\frac{C_4+C_8-2C_{10}}{\phi'}\right)\,,\\
{\cal E}&=&\left[C_1+\frac{1}{r^2}\left(\frac{C_3}{2h}-\frac{C_5}{\phi'}\right)\right]
\left(\phi''+\frac{\phi'h'}{2h}\right)
+\left[\frac{\phi'}{2}C_2-hC_4+\frac{1}{2r}\left(\frac{\phi'}{2}C_3-hC_5\right)\right]\frac{f'}{rf}
\notag\\
&&
+C_6+\frac{1}{r^2}\left(\frac{\phi'}{2}C_2-hC_4+C_8-2C_{10}\right)\,.
\ea
Varying the action (\ref{action}) with respect to $A_0$, 
it follows that  
\be
{\cal E}_{A_0} \equiv 
\left( G_{2,F} \sqrt{\frac{h}{f}}r^2 A_0' \right)'=0\,,
\label{EA}
\ee
whose integrated solution is given by 
\be
A_0'=\frac{1}{G_{2,F}} \sqrt{\frac{f}{h}}
\frac{q_0}{r^2}\,,
\label{A0dso}
\ee
where $q_0$ is constant. 
We will only focus on the case of an electric charge $q_0$
without considering the magnetic charge.

On using Eqs.~(\ref{back1}), (\ref{back2}), (\ref{back3}), 
and (\ref{EA}), 
the scalar field Eq.~(\ref{Ephi}) can be expressed as 
\be
{\cal E}_{\phi} \equiv -\frac{2}{\phi'}\left[\frac{f'}{2f}{\cal E}_{00}+{\cal E}_{11}'
+\left(\frac{f'}{2f}+\frac{2}{r}\right){\cal E}_{11}
+\frac{2}{r}{\cal E}_{22}-\frac{A_0'}{r^2} \sqrt{\frac{h}{f}}
{\cal E}_{A_0} \right]=0\,.
\label{Ephid}
\ee
We note that some of the coefficients appearing in the second-order 
action of even-parity perturbations derived later can be expressed 
in terms of the 
partial $\phi$ derivatives of ${\cal E}_{\phi}$ and ${\cal E}_{11}$.

\section{Odd-parity perturbations}
\label{oddsec}

On top of the static and spherically symmetric background 
(\ref{metric}), we decompose the metric tensor into the background 
and perturbed parts as $g_{\mu \nu}=\bar{g}_{\mu \nu}+h_{\mu \nu}$, 
where a bar represents the background quantity. 
Under the rotation in the $(\theta, \varphi)$ plane, the 
metric perturbations $h_{\mu \nu}$ can be separated into 
odd- and even-parity modes \cite{Regge:1957td,Zerilli:1970se}. 
Expanding $h_{\mu \nu}$ in terms of the spherical harmonics 
$Y_{l m} (\theta, \varphi)$, the odd- and even-modes of perturbations have 
parities $(-1)^{l+1}$ and $(-1)^l$, respectively. 
In the odd-parity sector, the components of $h_{\mu \nu}$ 
are given by 
\ba
& &
h_{tt}=h_{tr}=h_{rr}=0\,,\nonumber \\
& &
h_{ta}=\sum_{l,m}Q(t,r)E_{ab}
\nabla^bY_{lm}(\theta,\varphi)\,,
\qquad
h_{ra}=\sum_{l,m}W(t,r)E_{ab} \nabla^bY_{lm}(\theta,\varphi)\,,\nonumber \\
& &
h_{ab}=
\frac{1}{2}\sum_{l,m}
U (t,r) \left[
E_{a}{}^c \nabla_c\nabla_b Y_{lm}(\theta,\varphi)
+ E_{b}{}^c \nabla_c\nabla_a Y_{lm}(\theta,\varphi)
\right]\,,
\ea
where $Q$, $W$, and $U$ are functions of $t$ and $r$, and 
the subscripts $a$ and $b$ denote  either $\theta$ 
or $\varphi$ \cite{DeFelice:2011ka,Motohashi:2011pw,Kobayashi:2012kh,Kase:2014baa}. 
In a formal sense, we should write subscripts $l$ and $m$ 
for the variables $Q$, $W$, and $U$, but we omit 
them for brevity.
We note that $E_{ab}$ is an antisymmetric tensor with nonvanishing components 
$E_{\theta \varphi}=-E_{\varphi \theta}=\sin \theta$. 
The scalar field $\phi$ does not have an odd-parity perturbation, 
so it is equivalent to the background value $\phi(r)$.
The vector field $A_{\mu}$ in the odd-parity sector 
has the following perturbed components
\be
\delta A_{t}=\delta A_{r}=0\,,\qquad 
\delta A_{a}=\sum_{l,m} \delta A(t,r)E_{ab}
\nabla^b Y_{lm}(\theta,\varphi)\,,
\ee
where $\delta A$ depends on $t$ and $r$.

Under a gauge transformation $x_{\mu} \to x_{\mu}+\xi_{\mu}$, 
where $\xi_t=0$, $\xi_r=0$, and $\xi_a=\sum_{l,m} \Lambda(t,r) 
E_{ab} \nabla^b Y_{lm} (\theta, \varphi)$, the metric perturbations 
transform as $Q \to Q+\dot{\Lambda}$, 
$W \to W+\Lambda'-2\Lambda/r$, and $U \to U+2\Lambda$, 
where a dot represents the derivative with respect to $t$.
In the following, we choose the gauge 
\be
U=0\,,
\ee
which fixes the scalar $\Lambda$ in $\xi_a$.

We expand the action (\ref{action}) up to quadratic 
order in odd-parity perturbations. 
For this purpose it is sufficient to focus on the axisymmetric 
modes of perturbations characterized by $m=0$, since 
the nonaxisymmetric modes with $m \neq 0$ can be restored under 
the suitable rotation by virtue of the spherical symmetry on the background \cite{deRham:2020ejn}. 
We perform the integrals with respect to $\theta$ 
and $\varphi$ by using the following properties
\be
\int_0^{2\pi}{\rm d} \varphi \int_0^{\pi} {\rm d} \theta\,
Y_{l0, \theta}^2 \sin \theta=L\,,\qquad
\int_0^{2\pi}{\rm d} \varphi \int_0^{\pi} {\rm d} \theta\, 
\left( \frac{Y_{l0, \theta}^2}{\sin \theta}+
Y_{l0, \theta \theta}^2 \sin \theta \right)=L^2\,,
\label{Ythere}
\ee
where 
\be
L \equiv l (l+1)\,.
\ee
We also exploit the background Eqs.~(\ref{back1}), 
(\ref{back2}), and (\ref{EA}) to eliminate 
the terms $G_2$, $G_{2,X}$, and $A_0''$.
After integrating the action 
${\cal S}$
with respect to $t$ and $r$, 
the second-order action can be expressed in the form 
\be
{\cal S}_{\rm odd}=\sum_{l} L \int 
\rd t \rd r\,{\cal L}_{\rm odd}\,,
\label{Sodd}
\ee
where 
\ba
{\cal L}_{\rm odd} &=& \frac{\sqrt{h}}{4\sqrt{f}} {\cal H} 
\left( \dot{W}-Q'+\frac{2Q}{r} \right)^2
-\frac{\sqrt{h}}{\sqrt{f}} G_{2,F} A_0' 
\left( \dot{W}-Q'+\frac{2Q}{r} \right) \delta A
+(L-2) \left( \frac{{\cal F}Q^2}{4r^2 \sqrt{fh}}
-\frac{\sqrt{fh}}{4r^2} {\cal G}W^2 \right)
\nonumber \\
&&+\frac{1}{2 \sqrt{fh}}G_{2,F} 
\left( \dot{\delta A}^2-fh \delta A'^2-\frac{Lf}{r^2} \delta A^2 \right)\,,
\label{Lodd}
\ea
with
\ba
{\cal H}&\equiv&2 G_4+2 h\phi'^2G_{4,X}-h\phi'^2G_{5,\phi}
-\frac{h^2 \phi'^3 G_{5,X}}{r}
\,,\label{cHdef}\\
{\cal F}&\equiv&2 G_4+h\phi'^2G_{5,\phi}-h\phi'^2  \left( \frac12 h' \phi'+h \phi'' \right) G_{5,X}\,,
\label{cFdef}\\
{\cal G}&\equiv&2 G_4+2 h\phi'^2G_{4,X}-h\phi'^2 \left( G_{5,\phi}+{\frac {f' h\phi' G_{5,X}}{2f}} \right) \,.
\label{cGdef}
\ea
\subsection{$l \geq 2$}

We first derive linear stability conditions for the multipoles $l \geq 2$. 
To identify the dynamical DOFs, it is convenient to 
consider the following Lagrangian 
\ba
{\cal L}_{\rm odd} &=& \frac{\sqrt{h}}{4\sqrt{f}} {\cal H} 
\left[ 2\chi \left( \dot{W}-Q'+\frac{2Q}{r}
-\frac{2G_{2,F}A_0'}{{\cal H}}\delta A \right)-\chi^2 
\right]-\frac{1}{{\cal H}}\frac{\sqrt{h}}{\sqrt{f}} G_{2,F}^2 A_0'^2 
\delta A^2
+(L-2) \left( \frac{{\cal F}Q^2}{4r^2 \sqrt{fh}}
-\frac{\sqrt{fh}}{4r^2} {\cal G}W^2 \right)\nonumber \\
&&
+\frac{1}{2 \sqrt{fh}}G_{2,F} 
\left( \dot{\delta A}^2-fh \delta A'^2-\frac{Lf}{r^2} \delta A^2 \right)\,,
\label{Lodd2}
\ea
where we introduced an auxiliary field $\chi$. Variation of the Lagrangian (\ref{Lodd2}) with respect to 
$\chi$ leads to
\be
\chi=\dot{W}-Q'+\frac{2Q}{r}
-\frac{2G_{2,F}A_0'}{{\cal H}}\delta A\,.
\label{chi}
\ee
Substituting Eq.~(\ref{chi}) into Eq.~(\ref{Lodd2}), we find that 
the Lagrangian (\ref{Lodd2}) is equivalent to (\ref{Lodd}).
Varying (\ref{Lodd2}) with respect to $W$ and $Q$, respectively, 
we obtain
\ba
& &
(L-2)f {\cal G}W+r^2{\cal H}\,\dot{\chi}=0\,,\label{oddcon1}\\
& &\left[ 2(L-2){\cal F}Q+4rh {\cal H} \chi
+2r^2 h {\cal H} \chi'+r^2 ({\cal H}h'+2h {\cal H}')\chi 
\right]f-r^2 f' h{\cal H} \chi=0\,.
\label{oddcon2}
\ea
We solve Eqs.~(\ref{oddcon1}) and (\ref{oddcon2}) for $W$ 
and $Q$, and take the $t$ and $r$ derivatives of $W$ 
and $Q$, respectively. 
Substituting them into Eq.~(\ref{Lodd2}) and integrating it 
by parts, the resulting 
second-order Lagrangian can be expressed in the form 
\be
{\cal L}_{\rm odd}=
\dot{\vec{\mathcal{X}}}^{t}{\bm K}\dot{\vec{\mathcal{X}}}
+\vec{\mathcal{X}}'^{t}{\bm G}\vec{\mathcal{X}}'
+\vec{\mathcal{X}}^{t}{\bm M}\vec{\mathcal{X}}\,,
\label{LM2}
\ee
where ${\bm {K,G,M}}$ are $2\times2$ symmetric matrices, and $\vec{\mathcal{X}}$ is a vector field defined by 
\be
\vec{\mathcal{X}}=\left( 
\begin{array}{c}
\chi\\
\delta A
\end{array}
\right) \,.
\ee
Note that $\chi$ and $\delta A$ correspond to the dynamical 
perturbations arising from the gravity and vector field 
sectors, respectively.
The nonvanishing components of ${\bm {K,G,M}}$ are
\ba
& &
K_{11}=\frac{r^2 \sqrt{h}\,{\cal H}^2}{4(L-2)f^{3/2}{\cal G}}\,,
\qquad
K_{22}=\frac{G_{2,F}}{2\sqrt{fh}}\,,
\qquad
G_{11}=-fh \frac{{\cal G}}{{\cal F}}K_{11}\,,\qquad 
G_{22}=-fh K_{22}\,,\nonumber \\
& &
M_{11}=-\frac{\sqrt{h}\,{\cal H}}{4(L-2)\sqrt{f}}
\left( L-2+\alpha_M'-\frac{2}{r} \alpha_M \right)
\,,\qquad 
M_{22}=-\frac{G_{2,F}(L f{\cal H}+2r^2 h A_0'^2 G_{2,F})}
{2r^2 {\cal H} \sqrt{fh}}\,,\nonumber \\
& &
M_{12}=-\frac{\sqrt{h} A_0' G_{2,F}}{2\sqrt{f}},
\ea
where 
\be
\alpha_M \equiv 
-\frac{r^2h{\cal H}}{\cal F}\left(\frac{{\cal H}'}{\cal H}-\frac{f'}{2f}
+\frac{h'}{2h}+\frac{2}{r}\right)\,.
\ee
The ghosts are absent under the conditions $K_{11}>0$ and 
$K_{22}>0$. These translate to 
\ba
&& {\cal G}>0\,,
\label{nogoodd1}\\
& & G_{2,F}>0\,,
\label{nogoodd2}
\ea
respectively, which correspond to the no-ghost conditions of 
gravitational and vector field perturbations 
in the odd-parity sector.

The perturbation equations of motion for $\chi$ and $\delta A$ 
follow by varying the Lagrangian (\ref{LM2}) with respect to these variables. For the propagation of 
$\chi$ and $\delta A$ along the radial direction, 
we assume solutions to the perturbation equations in the form 
$\vec{\mathcal{X}}^{t} \propto e^{i (\omega t-kr)}$. 
In the short-wavelength limit $k \to \infty$, 
the dispersion relation is given by 
${\rm det} \left( \omega^2 {\bm K}+k^2{\bm G} 
\right)=0$. The radial propagation speed $c_r$ 
in proper time can be obtained by substituting $\omega=\sqrt{fh}\,c_r k$ 
into the dispersion relation. 
This gives the following two solutions
\ba
c_{r1,{\rm odd}}^2
&=& -\frac{G_{11}}{fh K_{11}}
=\frac{{\cal G}}{{\cal F}}\,,
\label{cr1}\\
c_{r2,{\rm odd}}^2
&=& -\frac{G_{22}}{fh K_{22}}
=1\,,
\label{cr2}
\ea
which are the squared propagation speeds of $\chi$ and $\delta A$, respectively. 
Under the no-ghost condition (\ref{nogoodd1}), 
the Laplacian stability of $\chi$ is ensured for 
\be
{\cal F}>0\,. 
\label{nolapodd1}
\ee
Since the second propagation speed squared (\ref{cr2}) is luminal, 
there is no Laplacian instability for $\delta A$.

In the large multipole limit $L=l(l+1) \gg 1$, the matrix ${\bm M}$ 
gives contributions to the propagation speed $c_{\Omega}$ 
along the angular direction. 
In this limit, we have 
\be
M_{11} \simeq -\frac{{\sqrt{h}\,\cal H}}{4\sqrt{f}} \,,\qquad 
M_{22} \simeq -\frac{\sqrt{f} G_{2,F}}{2r^2 \sqrt{h}}L\,.
\ee
Substituting solutions of the form 
$\vec{\mathcal{X}}^{t} \propto e^{i (\omega t-l \theta)}$ 
into the perturbation equations,   
the dispersion relation yields 
${\rm det}(\omega^2{\bm K}+{\bm M})=0$.
The angular propagation speed in proper time is given by 
$c_{\Omega}=\hat{c}_{\Omega}/\sqrt{f}$, 
where $\hat{c}_{\Omega}=r\,\rd \theta /\rd t$.
We substitute $\omega^2=\hat{c}_{\Omega}^2 l^2/r^2=
c_{\Omega}^2f l^2/r^2$ into the dispersion relation 
and solve it for $c_{\Omega}^2$. 
In the limit $l \gg 1$, we obtain the following two solutions
\ba
c_{\Omega 1,{\rm odd}}^2&=&
-\frac{r^2 M_{11}}{l^2 f K_{11}}
=\frac{{\cal G}}{{\cal H}}\,,
\label{co1}\\
c_{\Omega 2,{\rm odd}}^2&=&
-\frac{r^2 M_{22}}{l^2 f K_{22}}
=1\,,
\label{co2}
\ea
which correspond to the squared angular propagation 
speeds of $\chi$ and $\delta A$, respectively.
Under the no-ghost condition (\ref{nogoodd1}), 
the Laplacian instability of $\chi$ is absent for
\be
{\cal H}>0\,.
\ee
The angular propagation speed of $\delta A$ is luminal, so 
there is no Laplacian instability. 

We note that the stability conditions of $\chi$ are identical to 
those derived in Ref.~\cite{Kobayashi:2012kh} without a 
vector field $A_{\mu}$. 
This means that the presence of $A_{\mu}$ coupled to 
the scalar field of the form $G_2(\phi,X,F)$ does not modify the odd-parity 
stability conditions in the gravitational sector.
The odd-parity perturbation of $A_{\mu}$ 
propagates luminally, without a ghost for $G_{2,F}>0$.

\subsection{$l=1$}

We also study the odd-parity stability of dipolar perturbations ($l=1$). 
Since $L=2$ in this case, terms proportional to $L-2$ 
in Eq.~(\ref{Lodd}) vanish. 
Moreover, the metric components $h_{ab}$ vanish identically 
and hence $U=0$. To fix the residual gauge DOF, 
we choose the gauge 
\be
W=0\,.
\ee
Varying the Lagrangian (\ref{Lodd}) with respect to $W$ and $Q$, 
and setting $W=0$ at the end, we obtain 
\ba
\dot{\cal E}=0\,,\qquad 
\left( r^2 {\cal E} \right)'=0\,,
\label{Eeq}
\ea
where 
\be
{\cal E}={\cal H} \sqrt{\frac{h}{f}} \left( Q' 
-\frac{2}{r}Q+\frac{2G_{2,F}A_0'}{{\cal H}}
\delta A \right)\,.
\ee
Integrating two differential equations in (\ref{Eeq}) leads to 
\be
Q' -\frac{2}{r}Q+\frac{2G_{2,F}A_0'}{{\cal H}}
\delta A=\frac{1}{\cal H} \sqrt{\frac{f}{h}} 
\frac{{\cal C}}{r^2}\,,
\ee
where ${\cal C}$ is a constant.
On using this relation to eliminate $Q'$ from 
the Lagrangian (\ref{Lodd}), it follows that 
\be
{\cal L}_{\rm odd}=\frac{1}{2\sqrt{fh}} 
\left[ G_{2,F} \dot{\delta A}^2-G_{2,F}fh \delta A'^2
-\frac{2G_{2,F}(f {\cal H}+r^2 h G_{2,F}A_0'^2)}{r^2 {\cal H}} 
\delta A^2+\frac{f{\cal C}^2}{2r^4{\cal H}} \right]\,.
\label{Ll=1}
\ee
Hence the propagating DOF is only the 
vector field perturbation $\delta A$.
The ghost is absent so long as the first term in the 
square bracket of Eq.~(\ref{Ll=1}) is positive, i.e., 
\be
G_{2,F}>0\,,
\ee
which is the same as the no-ghost condition of $\delta A$ 
derived for $l \geq 2$.
In the short-wavelength limit, the dominant contributions to 
Eq.~(\ref{Ll=1}) are the first and second terms in the square bracket.
Then, the radial propagation speed squared of $\delta A$ 
in proper time is given by 
\be
c_{r,{\rm odd}}^2=1\,,
\ee
which is luminal. 
Thus, the stability of dipolar perturbations does not add any 
new conditions to those obtained for $l \geq 2$.

\section{Even-parity perturbations}
\label{evensec}

In this section, we derive the second-order action and perturbation equations of motion for the even-parity modes.  
On the background (\ref{metric}), the metric perturbations $h_{\mu \nu}$ in the even-parity sector are 
given by
\ba
&&
h_{tt}=f(r) \sum_{l,m}H_0(t,r)Y_{lm}(\theta,\varphi)\,,\qquad
h_{tr}=h_{rt}=\sum_{l,m}H_1(t,r)Y_{lm}(\theta,\varphi)\,,\qquad
h_{rr}=h(r)^{-1}\,\sum_{l,m}H_2(t,r)Y_{lm}(\theta,\varphi)\,,\notag\\
&&
h_{ta}=h_{at}=\sum_{l,m}h_0(t,r)\nabla_aY_{lm}(\theta,\varphi)\,,\qquad
h_{ra}=h_{ar}=\sum_{l,m}h_1(t,r)\nabla_aY_{lm}(\theta,\varphi)\,,\notag\\
&&
h_{ab}=\sum_{l,m}\left[K(t,r)g_{ab}Y_{lm}(\theta,\varphi)
+G(t,r)\nabla_a\nabla_bY_{lm}(\theta,\varphi)\right]\,, 
\label{evenmetric}
\ea
where $H_0$, $H_1$, $H_2$, $h_0$, $h_1$, $K$, and $G$ are scalar quantities 
depending on $t$ and $r$. 
We also decompose the scalar and vector fields as 
\ba
\phi&=&\bar{\phi}(r)+\sum_{l,m}\delta\phi(t,r)Y_{lm}(\theta,\varphi)\,,\label{defdphi}\\
A_{\mu}&=&\bar{A}_{\mu}+\delta A_{\mu}\,,\label{defdA}
\ea
with
\be
\delta A_t=\sum_{l,m}\delta A_0(t,r)Y_{lm}(\theta, \varphi)\,,\qquad
\delta A_r=\sum_{l,m}\delta A_1(t,r)Y_{lm}(\theta, \varphi)\,,\qquad 
\delta A_a=\sum_{l,m}\delta A_2(t,r)\nabla_{a}Y_{lm}(\theta, \varphi)\,, 
\label{vecpert}
\ee
where $\delta\phi$, $\delta A_0$, $\delta A_1$, and $\delta A_2$ are 
functions of $t$ and $r$. 

Under the infinitesimal gauge transformation $x_\mu\to x_\mu+\xi_\mu$ with 
\be
\xi_t=\sum_{l,m} {\cal T}(t,r)Y_{lm}(\theta,\varphi)\,,\qquad 
\xi_r=\sum_{l,m} {\cal R}(t,r)Y_{lm}(\theta,\varphi)\,,\qquad 
\xi_a=\sum_{l,m} \Theta(t,r) \nabla_a Y_{lm}(\theta,\varphi)\,,
\label{xi}
\ee
the metric perturbations in Eq.~(\ref{evenmetric}) and $\delta \phi$ 
in Eq.~(\ref{defdphi}) transform as 
\ba
& &
H_0 \to H_0+\frac{2}{f} \dot{{\cal T}}-\frac{f' h}{f}{\cal R}\,,\qquad 
H_1 \to H_1+\dot{{\cal R}}+{\cal T}'-\frac{f'}{f}{\cal T}\,,\qquad
H_2 \to H_2+2h{\cal R}'+h' {\cal R}\,,\label{H1tra} \notag \\
& &
h_0 \to h_0+{\cal T}+\dot{\Theta}\,,\qquad 
h_1 \to h_1+{\cal R}+\Theta'
-\frac{2}{r}\Theta\,,\qquad
K \to K+\frac{2}{r}h{\cal R}\,,\qquad 
G \to G+\frac{2}{r^2}\Theta\,,\notag\\
&&
\delta\phi\to\delta\phi-\phi'h{\cal R}\,.
\label{gaugetrans}
\ea
We can eliminate some of the perturbed variables on account  
of the gauge DOFs. 
For the multipoles $l\geq2$, we choose the uniform 
curvature gauge given by 
\be
h_0=0\,,\qquad
G=0\,,\qquad
K=0\,,
\label{gauge1}
\ee
under which ${\cal R}$, $\Theta$, and ${\cal T}$ are fixed.
In addition to the coordinate transformation (\ref{xi}), the action (\ref{action}) is 
invariant under the $U(1)$ gauge transformation
\be
\delta A_{\mu} \to \delta A_{\mu}+\partial_{\mu} \delta \chi
\qquad {\rm with} \qquad
\delta \chi=\sum_{l,m} \tilde{\chi}(t,r)Y_{lm}(\theta,\varphi)\,. 
\ee
Under this transformation, the scalar quantities of vector field perturbations 
in Eq.~(\ref{vecpert}) transform as 
\be
\delta A_0\to\delta A_0+\dot{\tilde{\chi}}\,,\qquad
\delta A_1\to\delta A_1+\tilde{\chi}'\,,\qquad
\delta A_2\to\delta A_2+\tilde{\chi}\,.
\ee
We choose the gauge 
\be
\delta A_2=0\,, 
\label{gauge2}
\ee
under which $\tilde{\chi}$ is fixed.

\subsection{Second-order action and perturbation equations of motion}

We expand the action (\ref{action}) up to second order with 
the gauge choices (\ref{gauge1}) and (\ref{gauge2}). 
As in the case of odd-parity modes, we set $m=0$ 
without loss of generality. 
Performing the integration by parts and using the background equations of motion 
(\ref{back1})-(\ref{back3}) 
and (\ref{EA}), the second-order action of even-parity 
perturbations can be expressed in the form 
\be
{\cal S}_{\rm even} = \sum_l\int {\rm d}t\, {\rm d}r 
\left({\cal L}_{u}+{\cal L}_{A}\right)\,,
\label{acteven}
\ee
where
\ba
{\cal L}_{u}
&=&
H_0 \left[ a_1 \delta \phi'' +a_2 \delta \phi' +a_3 H_2'
+L a_4 h_1'+\left( a_5+L a_6 \right) \delta \phi 
+\left( a_7+L a_8 \right) H_2+L a_9 h_1 \right] 
\notag \\
&&
+L b_1 H_1^2
+H_1 \left( b_2 \dot{\delta \phi}'+b_3 \dot{\delta \phi}
+b_4 \dot{H}_2+L b_5 \dot{h}_1 \right)
+c_1 \dot{\delta \phi} \dot{H}_2
+H_2 \left[ c_2 \delta \phi'+ (c_3+L c_4) \delta \phi
+L c_5 h_1 \right]
\notag \\
&&
+c_6 H_2^2 
+L d_1 \dot{h}_1^2
+L h_1 \left( d_2 \delta \phi'
+d_3 \delta \phi \right)+L d_4 h_1^2
+e_1 \dot{\delta \phi}^2+e_2 \delta \phi'^2
+\left( e_3+L e_4 \right) \delta \phi^2
\,,
\label{Lu}
\\
{\cal L}_{A}
&=&
v_1(\delta A_0'-\dot{\delta A}_1)^2
+(\delta A_0'-\dot{\delta A}_1)
\left(v_2H_0+v_3H_2+v_4\delta\phi'+v_5\delta\phi\right)
+v_6H_0^2
\notag \\
&&
+L(v_7\delta A_0 h_1+v_8\delta A_0^2+v_9\delta A_1^2)\,.
\label{LdA}
\ea
We recall that $L$ is defined by $L=l(l+1)$.
The coefficients $a_1$, $a_2$, ..., $v_9$ are given in Appendix~\ref{AppA}. 
In comparison to Horndeski theories without the Maxwell field, 
the Lagrangian ${\cal L}_u$ has a same structure with 
that derived in Refs.~\cite{Kobayashi:2014wsa,Kase:2021mix}. 
Still, the coefficients $a_2$, $a_5$, $a_7$, $b_3$, $c_2$, $c_3$, 
$c_6$, $e_1$, $e_2$ are subject to modifications by 
the presence of $A_{\mu}$ (see Appendix~\ref{AppA}). 
Moreover, the vector field perturbation gives rise to the new Lagrangian 
(\ref{LdA}) whose contribution is absent in Refs.~\cite{Kobayashi:2014wsa,Kase:2021mix}. 

In what follows, we derive the perturbation equations of motion 
by varying 
the second-order action (\ref{acteven}) with respect to  
$H_0$, $H_1$, $H_2$, $h_1$, $\delta A_0$, $\delta A_1$, $\delta\phi$, 
and eliminate nondynamical variables from the action 
by using their corresponding equations. The Lagrangian (\ref{LdA}) shows that $\delta A_0$ is 
nondynamical since there is no quadratic term of its time derivative. Nevertheless, the perturbation equation of $\delta A_0$ cannot be explicitly solved for $\delta A_0$ due to the existence of the 
quadratic radial derivative term, i.e., $\delta A_0'^2$, in Eq.~(\ref{LdA}). 
This situation is similar to the case of odd-parity perturbations 
discussed in Sec.~\ref{oddsec}. 
Thus, we introduce an auxiliary field $V(t,r)$ in analogy to 
the discussion in the odd-parity sector, 
and rewrite Eq.~(\ref{LdA}) in the form 
\ba
{\cal L}_{A}
&=&
v_1\left[2V\left(\delta A_0'-\dot{\delta A}_1+
\frac{v_2H_0+v_3H_2+v_4\delta\phi'+v_5\delta\phi}{2v_1}\right)
-V^2\right]
-\frac{\left(v_2H_0+v_3H_2+v_4\delta\phi'+v_5\delta\phi\right)^2}{4v_1}
\notag\\
&&
+v_6H_0^2+L(v_7\delta A_0 h_1+v_8\delta A_0^2+v_9\delta A_1^2)\,.
\label{LdA2}
\ea
Varying this action with respect to $V$ gives
\be
V=\delta A_0'-\dot{\delta A}_1+
\frac{v_2H_0+v_3H_2+v_4\delta\phi'+v_5\delta\phi}{2v_1}\,.
\label{Vexp}
\ee
Substituting Eq.~(\ref{Vexp}) into Eq.~(\ref{LdA2}), we find that Eq.~(\ref{LdA2}) 
is equivalent to the original Lagrangian (\ref{LdA}). 
By introducing the auxiliary field $V$, the quadratic terms 
$\delta A_0'^2$ and $\dot{\delta A}_1^2$ present in the original 
Lagrangian (\ref{LdA}) are absent in the new Lagrangian (\ref{LdA2}). 
This allows us to solve the perturbation equations of $\delta A_0$ and $\delta A_1$ explicitly for themselves. 
Substituting such solutions into the Lagrangian, we will see later that the dynamical property of vector field perturbation is determined by the auxiliary field $V$.

We also note that the coefficients $v_2$ and $v_6$ 
have the following relation 
\be
v_6=\frac{v_2^2}{4v_1}\,.
\label{v6v2}
\ee
This means that the two quadratic terms of $H_0$ appearing 
in Eq.~(\ref{LdA2}), i.e., $-[v_2^2/(4v_1)]H_0^2$ and $v_6 H_0^2$ 
cancel each other as a result of introducing the auxiliary field $V$.
Consequently, the total action (\ref{acteven}) with the sum of 
Eqs.~(\ref{Lu}) and (\ref{LdA2}) depends on $H_0$ linearly. 
Hence the perturbation $H_0$ corresponds to a Lagrange multiplier and
the variation of the action with respect to $H_0$
puts constraint on other perturbation variables.

Varying the total action (\ref{acteven}) with Eqs.~(\ref{Lu}) and (\ref{LdA2}) 
with respect to $H_0$, $H_1$, $H_2$, $h_1$, $\delta A_0$, $\delta A_1$, 
and $\delta\phi$, we obtain the following linear 
perturbation equations
\ba
&&
a_1\delta\phi''+a_3H_2'+La_4h_1'+\left(a_2-\frac{v_2v_4}{2v_1}\right)\delta\phi'
+\left(a_5+La_6-\frac{v_2v_5}{2v_1}\right)\delta\phi
+\left(a_7+La_8-\frac{v_2v_3}{2v_1}\right)H_2
\notag\\
&&
+La_9h_1
+v_2V=0\,,
\label{eqH0}\\
&&
2Lb_1H_1+b_2\delta\dot{\phi}'+b_3\delta\dot{\phi}+b_4\dot{H}_2+Lb_5\dot{h}_1=0\,,
\label{eqH1}\\
&&
-c_1 \delta\ddot{\phi} -b_4\dot{H}_1 
+\left(c_2-\frac{v_3v_4}{2v_1}\right) \delta\phi' + \left(c_3+Lc_4-\frac{v_3v_5}{2v_1}\right) \delta\phi+Lc_5 h_1
+\left(2 c_6-\frac{v_3^2}{2v_1}\right) H_2
-a_3H_0'
\notag\\
&&
+ \left( a_7-a_3'+La_8-\frac{v_2v_3}{2v_1} \right) H_0+v_3V=0\,,
\label{eqH2}\\
&&
-2 d_1 \ddot{h}_1+d_2 \delta\phi'+d_3 \delta\phi+2 d_4 h_1
-a_4H_0' +( a_9-a_4' ) H_0-b_5\dot{H}_1+ c_5 H_2
+v_7\delta A_0=0\,,
\label{eqh1}\\
&&
-2(v_1V)'+Lv_7h_1+2Lv_8\delta A_0=0\,,
\label{eqdA0}\\
&&
2v_1\dot{V}+2Lv_9\delta A_1=0\,,
\label{eqdA1}\\
&&
-2 e_1 \ddot{\delta\phi}-\left(2 e_2-\frac{v_4^2}{2v_1}\right) \delta\phi''
+ \left[2 e_3+2Le_4 +\left(\frac{v_4v_5}{2v_1}\right)'-\frac{v_5^2}{2v_1}\right] \delta\phi
+a_1H_0'' + \left( 2 a_1'-a_2 +\frac{v_2v_4}{2v_1}\right) H_0'
\notag\\
&&
+ \left[ a_1''-a_2'+a_5+La_6+\left(\frac{v_2v_4}{2v_1}\right)'-\frac{v_2v_5}{2v_1} \right] H_0
+b_2\dot{H}_1' + ( b_2'-b_3 ) \dot{H}_1-c_1 \ddot{H}_2-\left(c_2-\frac{v_3v_4}{2v_1}\right)H_2' 
\notag\\
&&
+ \left[ c_3-c_2'+Lc_4 +\left(\frac{v_3v_4}{2v_1}\right)'-\frac{v_3v_5}{2v_1}\right] H_2
-Ld_2h_1' +L ( d_3-d_2' ) h_1-\left[2 e_2'-\left(\frac{v_4^2}{2v_1}\right)' \right]\delta\phi'
\notag\\
&&
-v_4V'+(v_5-v_4')V =0\,,
\label{eqdphi}
\ea
where we used the relation (\ref{v6v2}).

\subsection{Linear stability conditions}

In order to derive the linear stability conditions of dynamical perturbations, 
we eliminate nondynamical variables from the total action (\ref{acteven}) 
with Eqs.~(\ref{Lu}) and (\ref{LdA2}) by using some of 
the equations derived above. Since the number of dynamical 
perturbations is different depending on the values of $l$, 
we investigate the three cases 
(1) $l\geq2$, (2) $l=0$, and (3) $l=1$, in turn.

\subsubsection{$l\geq2$}

Among the eight variables 
($H_0$, $H_1$, $H_2$, $h_1$, $\delta A_0$, $\delta A_1$, $\delta\phi$, $V$), 
we can eliminate $H_1$, $\delta A_0$, and $\delta A_1$ by using 
Eqs.~(\ref{eqH1}), (\ref{eqdA0}), and (\ref{eqdA1}), respectively. 
This is due to the fact that the derivatives of $H_1$, $\delta A_0$, 
and $\delta A_1$ do not appear in their equations.
We recall that $H_0$ corresponds to a Lagrange multiplier, so 
its perturbation equation (\ref{eqH0}) puts constraint on 
other variables. 
Introducing a new variable~\cite{Kobayashi:2014wsa,Kase:2021mix}
\be
\psi=H_2+\frac{a_4}{a_3}Lh_1+\frac{a_1}{a_3}\delta\phi'\,,
\label{defpsi}
\ee
we can write Eq.~(\ref{eqH0}) in terms of 
$\psi'$, $\psi$, $\delta \phi'$, $\delta \phi$, $V$, and $h_1$. 
We solve this equation for $h_1$ and take its time derivative.
Terms $H_2$ and $\dot{H}_2$ in the action (\ref{acteven}) can be expressed 
in terms of $\psi$, $\dot{\psi}$, $\delta\phi'$, $\dot{\delta\phi}'$, $h_1$, 
$\dot{h}_1$, where the latter two variables now depend on $\psi$, $\delta \phi$, 
$V$ and their derivatives.
Then, we can express the 
second-order action (\ref{acteven}) 
in terms of the three dynamical perturbations $\psi$, $\delta\phi$, $V$ and their derivatives. 
After the integration by parts, we obtain
\be
{\cal S}_{\rm even}  = \sum_l\int {\rm d}t\, {\rm d}r 
\left(\dot{\vec{\mathcal{X}}}^{t}{\bm K}\dot{\vec{\mathcal{X}}}
+\vec{\mathcal{X}}'^{t}{\bm G}\vec{\mathcal{X}}'
+\vec{\mathcal{X}}^{t}{\bm Q}\vec{\mathcal{X}}'
+\vec{\mathcal{X}}^{t}{\bm M} \vec{\mathcal{X}}\right)\,, 
\label{evenact}
\ee
where ${\bm K}$, ${\bm G}$, ${\bm M}$ are the $3\times3$ symmetric matrices while ${\bm Q}$ is antisymmetric, and the vector $\vec{\mathcal{X}}$ is defined as 
\be
\vec{\mathcal{X}}=\left( 
\begin{array}{c}
\psi\\
\delta \phi\\ 
V
\end{array}
\right) \,.
\label{calX}
\ee
Note that the derivative terms $\dot{\delta\phi}'$ and $\dot{\psi}'$ disappear from the final action (\ref{evenact}).

The kinetic matrix ${\bm K}$ in the reduced action (\ref{evenact}) must be positive definite 
for the absence of ghosts. This requires that the determinants of 
principal submatrices of ${\bm K}$ are positive, such that 
\ba
K_{33}
&=&
\frac{v_1^2}{Lfh v_8}=\frac{2v_1^2}{L\sqrt{fh}\,G_{2,F}}
>0\,,
\label{K33}\\
K_{11}K_{33}-K_{13}^2
&=&
\frac{(L{\cal P}_1-{\cal F})f^3{\cal P}_2^4 v_1^2}
{L^2 h^3 (rf'-2f)^4{\cal H}^2 ({\cal P}_2+2rL{\cal H})^2 G_{2,F}}
>0\,,
\label{detK1}\\
{\rm det}{\bm K}
&=&
\frac{(L-2) f^{5/2} {\cal F}v_1^2 {\cal P}_2^4 
(2{\cal P}_1-{\cal F})}{2L^2h^{7/2}{\cal H}^2 \phi'^2 
({\cal P}_2+2rL{\cal H})^2 (rf'-2f)^4 G_{2,F}}
>0\,,\label{detK2}
\ea
where we introduced the following combinations~\cite{Kobayashi:2014wsa}
\ba
{\cal P}_1 \equiv \frac{h \mu}{2fr^2 {\cal H}^2} \left( 
\frac{fr^4 {\cal H}^4}{\mu^2 h} \right)'\,,\qquad 
{\cal P}_2 \equiv \frac{h}{f}(rf'-2f)\mu
\,,\qquad 
\mu\equiv\frac{2(\phi' a_1+2ra_4)}{\sqrt{fh}}\,.
\label{defP1P2}
\ea
Under the stability conditions (\ref{nogoodd2}) and (\ref{nolapodd1}) 
in the odd-parity sector, the quantities $G_{2,F}$ and ${\cal F}$ are positive. 
Then, the first condition (\ref{K33}) is automatically satisfied. 
Remembering that $L>2$, both the second and third inequalities  
(\ref{detK1})-(\ref{detK2}) hold for 
\be
{\cal K} \equiv 2{\cal P}_1-{\cal F}>0\,, 
\label{nogoeven}
\ee
which coincides with the stability condition in Horndeski theories 
without the Maxwell field~\cite{Kobayashi:2014wsa}. 
Consequently, the absence of ghost instabilities in the even-parity 
sector adds one condition (\ref{nogoeven}) to 
the stability conditions in the odd-parity sector. 

We proceed to derive the propagation speeds of even-parity perturbations 
along the radial direction. 
The equations of motion for three dynamical perturbations 
follow by varying the action (\ref{evenact}) 
with respect to $\vec{\mathcal{X}}$.
Assuming the solutions to those equations of the form 
$\vec{\mathcal{X}}\propto e^{i( \omega t-kr)}$,  
where $\omega$ and $k$ are the frequency and wavenumber, 
respectively, we obtain the dispersion relation characterizing the 
radial propagation. 
In the limits of large $\omega$ and $k$, it is given by 
\be
{\rm det}|fhc_r^2{\bm K}+{\bm G}|=0\,. 
\label{dispr}
\ee
Here, the propagation speed $c_r$ is defined by the rescaled radial coordinate 
$r_*=\int {\rm d}r/\sqrt{h}$ and proper time $\tau=\int \sqrt{f}\,{\rm d}t$, 
as $c_r={\rm d}r_*/{\rm d}\tau=(fh)^{-1/2}({\rm d}r/{\rm d}t)=(fh)^{-1/2}(\omega/k)$. 
The matrix components of ${\bm K}$ and ${\bm G}$ associated with 
the vector field perturbation $V$ have the following relations, 
\be
\frac{G_{13}}{K_{13}}=
\frac{G_{23}}{K_{23}}=
\frac{G_{33}}{K_{33}}=-fh\,.
\ee
On using these relations,  the radial propagation speed of $V$, which 
is decoupled from the other two, is simply given by 
\be
c_{r3,{\rm even}}^2=1\,,
\label{cr3even}
\ee
which is equivalent to the radial propagation speed of vector field 
perturbation $\delta A$ in the odd-parity sector (\ref{cr2}). 

The other components of matrices ${\bm K}$ and ${\bm G}$ 
are quite complicated, but we can resort to the following relation 
\be
a_4'=\frac {1}{2 f-rf'} \left[ \left( rf''-\frac {r{f'}^{2}}{f}+2 f'-\frac {2f}{r} \right) a_4
+\frac {{f}^{3/2}}{r\sqrt {h}} {\cal F} -2rfhA_0'^2v_8 \right] \,, 
\label{conda4}
\ee
to eliminate the derivative $a_4'$. This relation follows by using the 
background Eqs.~(\ref{back1}) and (\ref{back3}). 
As a consequence, the dispersion relation can be 
factorized in the form, 
\be
(c_r^2-c_{r1,{\rm even}}^2)(c_r^2-c_{r2,{\rm even}}^2)=0\,, 
\ee
where $c_{r1,{\rm even}}$ and $c_{r2,{\rm even}}$ correspond
to the radial propagation speeds of $\psi$ and $\delta \phi$, respectively, 
which are given by 
\ba
c_{r1,{\rm even}}^2 &=&
\frac{\cal G}{\cal F}\,,
\label{cr1even}\\
c_{r2,{\rm even}}^2 &=&
\frac { 4\phi' }{(fh)^{3/2}(2{\cal P}_1-{\cal F}) \mu^2}
\left[ 8r^2 h a_4 c_4 (\phi' a_1+ra_4)
-\sqrt{fh}\phi'a_1^2 {\cal G} +2r^2a_4^2 \left( \frac{f'}{f}a_1+2 c_2
+A_0'v_4+\frac{\phi'v_4^2}{2v_1} \right) \right]
\,.\qquad
\label{cr2even}
\ea
Notice that $c_{r1,{\rm even}}^2$ is equivalent to the squared propagation 
speed (\ref{cr1}) of gravitational perturbation $\chi$ in the odd-parity sector, 
which is not directly affected by the presence of the vector field.
On the other hand, the coupling between $\phi$ and $A_{\mu}$
modifies the value of $c_{r2,{\rm even}}$ due to the presence of 
the last two terms in Eq.~(\ref{cr2even}) containing 
$v_4=-{r^2h^{3/2}\phi'A_0'G_{2,XF}}/{\sqrt{f}}$. 
In the absence of the vector field, the result (\ref{cr2even}) 
is consistent with those derived in 
Refs.~\cite{Kobayashi:2014wsa,Kase:2021mix}. 

We will also obtain the propagation speeds of even-parity perturbations 
along the angular direction. 
Fo this purpose, we assume the solution to the equations of dynamical 
perturbations in the form $\vec{\mathcal{X}}\propto e^{i(\omega t-l \theta)}$. 
In the limit of large $\omega$ and $l$, the reduced Lagrangian (\ref{evenact}) leads to the following dispersion relation along the angular direction 
\be
{\rm det}|fl^2c_{\Omega}^2{\bm K}+r^2{\bm M}|=0\,.
\label{dispo}
\ee
The propagation speed $c_{\Omega}$ is defined by using the 
proper time $\tau$ such that 
$c_{\Omega}=r {\rm d}\theta/{\rm d}\tau=(r/\sqrt{f})({\rm d}\theta/{\rm d}t)
=(r/\sqrt{f})(\omega/l)$. Expanding the components of ${\bm K}$ and ${\bm M}$ in the limit $l\to\infty$, we find that the leading-order 
matrix components have the following 
dependence\footnote{Nonvanishing components of 
the antisymmetric matrix ${\bm Q}$ have the following 
leading-order $l$-dependence 
\be
Q_{12}=\frac{\tilde{Q}_{12}}{l^{2}}\,,\qquad
Q_{13}=\frac{\tilde{Q}_{13}}{l^{2}}\,,\qquad
Q_{23}=\frac{\tilde{Q}_{23}}{l^{2}}\,.
\ee
These do not contribute to the values of $c_{\Omega}$ 
in the large $l$ limit.
}, 
\ba
&&
K_{11}=\frac{\tilde{K}_{11}}{l^{4}}\,,\qquad
K_{12}=\frac{\tilde{K}_{12}}{l^{2}}\,,\qquad
K_{13}=\frac{\tilde{K}_{13}}{l^{4}}\,,\qquad
K_{22}=\tilde{K}_{22}\,,\qquad
K_{23}=\frac{\tilde{K}_{23}}{l^{2}}\,,\qquad
K_{33}=\frac{\tilde{K}_{33}}{l^{2}}\,,\nonumber \\
&&
M_{11}=\frac{\tilde{M}_{11}}{l^{2}}\,,\qquad
M_{12}=\tilde{M}_{12}\,,\qquad
M_{13}=\frac{\tilde{M}_{13}}{l^{2}}\,,\qquad
M_{22}=\tilde{M}_{22}l^2\,,\qquad
M_{23}=\tilde{M}_{23}\,,\qquad
M_{33}=\tilde{M}_{33}\,,
\ea
where the quantities with tildes do not contain the $l$ dependence. 
Picking up the leading-order contributions to Eq.~(\ref{dispo}), 
it follows that 
\be
(f\tilde{K}_{33}c_{\Omega}^2+r^2\tilde{M}_{33})
\left[
f^2(\tilde{K}_{11}\tilde{K}_{22}-\tilde{K}_{12}^2)c_{\Omega}^4
+r^2f(\tilde{K}_{11}\tilde{M}_{22}-2\tilde{K}_{12}\tilde{M}_{12}+\tilde{K}_{22}\tilde{M}_{11})c_{\Omega}^2
+r^4(\tilde{M}_{11}\tilde{M}_{22}-\tilde{M}_{12}^2)\right]=0\,.
\label{dispo2}
\ee
Then, the propagation speed of $V$ decouples from 
the other two, such that 
\be
c_{\Omega3,{\rm even}}^2=-\frac{r^2\tilde{M}_{33}}{f\tilde{K}_{33}}
=\frac{r^2hv_8}{v_1}=\frac{G_{2,F}}{G_{2,F}+2F G_{2,FF}}\,,
\label{cO3}
\ee
where $F$ is given in Eq.~(\ref{BGF}).
If the Lagrangian $G_2$ contains nonlinear functions of $F$, 
the propagation speed of $V$ along the angular direction 
deviates from that of light. This property dose not necessarily 
hold in other spacetime since the propagation speed of perturbations generally depends on underlying symmetry of the background spacetime. 
Indeed, on the Friedmann-Lema\^{i}tre-Robertson-Walker (FLRW) cosmological background, the propagation speeds of vector perturbations are luminal 
in theories with the coupling $G_2=G_2(F)$ \cite{Kase:2018nwt}.

Under the no-ghost condition $G_{2,F}>0$, the angular Laplacian stability of 
$V$ is ensured for 
\be
G_{2,F}+2F G_{2,FF}>0\,.
\ee
The other two propagation speeds associated with the perturbations 
$\psi$ and $\delta \phi$ are given by 
\be
c_{\Omega\pm,{\rm even}}^2=-B_1\pm\sqrt{B_1^2-B_2}\,, 
\label{cosqeven}
\ee
where 
\be
B_1=\frac{r^2(\tilde{K}_{11}\tilde{M}_{22}
-2\tilde{K}_{12}\tilde{M}_{12}+\tilde{K}_{22}\tilde{M}_{11})}
{2f (\tilde{K}_{11}\tilde{K}_{22}-\tilde{K}_{12}^2)}\,,\qquad 
B_2=\frac{r^4(\tilde{M}_{11}\tilde{M}_{22}-\tilde{M}_{12}^2)}
{f^2 (\tilde{K}_{11}\tilde{K}_{22}-\tilde{K}_{12}^2)}\,.\label{defB1B2}
\ee
While each matrix component of ${\bm K}$ and ${\bm M}$ is quite complicated, 
we can simplify the terms appearing in Eq.~(\ref{defB1B2}) by using 
relations among the coefficients given in Appendix~\ref{AppA}. 
We also exploit the following relation
\ba
&&
(2h\phi''+h'\phi')\left[\frac{a_1-r^2hc_4}{r^2\sqrt{fh}}+\frac{f({\cal G}-{\cal H})}{rh(rf'-2f)\phi'}\right] 
-\frac{h}{4f}\left[2f''-\frac{(rf'-2f)f'}{rf}+\frac{(rf'-4f)h'}{rh}\right]{\cal H}\notag\\
&&
-\frac{1}{r^2}{\cal F}-\frac{rh'-2h}{2r^2}{\cal G}-\frac{h(rf'-2f)\phi'}{2rf}\frac{\partial {\cal H}}{\partial\phi}
+\frac{2h^{3/2}A_0'^2}{\sqrt{f}}v_8=0\,,
\ea
which is equivalent to the subtraction of Eq.~(\ref{back1}) from Eq.~(\ref{back3}). 
After lengthy but straightforward calculations, we find that the quantities $B_1$ and $B_2$ are of 
the same forms as those derived in Ref.~\cite{Kase:2021mix} 
without a perfect fluid, i.e., 
\ba
\hspace{-0.8cm}
&&
B_1=
{\frac { a_4 r^3 [ 4 h ( \phi' a_1+2 ra_4 ) \beta_1+\beta_2
-4 \phi' a_1 \beta_3  ] 
-2 fh{\cal G}  [ 2 ra_4  ( 2 {\cal P}_1-{\cal F} )  ( \phi' a_1+ra_4 ) 
+\phi'^2a_1^{2}{\cal P}_1 ] }
{ 4\sqrt {fh} a_4( \phi' a_1+2 ra_4 )^2 ( 2 {\cal P}_1-{\cal F} )}},
\label{B1def}\\
&&
B_2=
-r^2{\frac {r^2h \beta_1 [ 2 fh {\cal F} {\cal G} ( \phi' a_1+2 ra_4 ) +r^2\beta_2 ] 
-{r}^{4}\beta_2 \beta_3
-fh{\cal F} {\cal G}  ( \phi' fh {\cal F} {\cal G}a_1 +4 r^3 a_4 \beta_3 ) }
{ fh{\cal F} \phi' a_1  ( \phi' a_1+2 ra_4 ) ^{2} ( 2 {\cal P}_1-{\cal F} )  }}\,,
\label{B2def}
\ea
with 
\ba
\beta_1&=&\phi'^2a_4 e_4-2 \phi' c_4 a_4'
+ \left[  \left( {\frac {f'}{f}}+{\frac {h'}{h}}-\frac{2}{r} \right) a_4
+{\frac {\sqrt {fh}{\cal F}}{r}} \right] \phi' c_4+{\frac {f{\cal F} {\cal G}}{2r^2}}\,,\\
\beta_2&=& \left[ {\frac {\sqrt {fh}{\cal F}}{r^2} 
\left( 2 hr\phi'^2c_4+{\frac {rf' \phi'a_4}{f}}-\sqrt {fh}\phi'{\cal G} \right) }
-{\frac {2\sqrt {fh}\phi' a_4 {\cal G}}{r} 
\left( {\frac {{\cal G}'}{{\cal G}}}-{\frac {a_4'}{a_4}}+{\frac {f'}{f}}+\frac12 {\frac {h'}{h}}-\frac{1}{r} \right) } \right] a_1-{\frac {4{\cal F} {\cal G} fha_4}{r}}\,,\qquad\,\,\\
\beta_3&=& \left( hc_4'-\frac{d_3}{2}+\frac12 h' c_4 \right) \phi' a_4
+ \left( {\frac {h'}{2h}}-\frac{1}{r}+{\frac {f'}{2f}}-{\frac {a_4'}{a_4}} \right)  
\left( {\frac {a_4 f'}{2f}}+2 h \phi'c_4+{\frac {\sqrt {fh}{\cal G}}{2r}} \right) a_4\notag\\
&&
+{\frac {\sqrt {fh}{\cal F}}{4r} \left( {\frac {f'}{f}}a_4+2 h \phi'c_4+{\frac {3\sqrt {fh}{\cal G}}{r}} \right) }\,.
\ea
While $B_1$ and $B_2$ do not explicitly contain the vector field contribution, 
the quantity $a_4'$ present in $\beta_1$, $\beta_2$, $\beta_3$ 
generally picks up such contributions, see Eq.~(\ref{conda4}). 
To ensure the Laplacian stabilities of perturbations $\psi$ 
and $\delta \phi$, we require that $c_{\Omega\pm,{\rm even}}^2>0$. 
These conditions are satisfied if 
\be
B_1^2 \geq B_2>0 \quad {\rm and} \quad 
B_1<0\,.
\ee
%

\vskip-\baselineskip
\begin{table}[ht]
\renewcommand\thetable{\!\!}
\newcommand\xrowht[2][0]{\addstackgap[.5\dimexpr#2\relax]{\vphantom{#1}}}
\newcolumntype{C}[1]{>{\hfil}m{#1}<{\hfil}}
\centering
\caption{Summary of linear stability conditions.}
\begin{tabular}{|C{30mm}|C{28mm}|C{25mm}|C{68mm}|} 
\hline\xrowht{12pt}
     & No ghosts & $c_r^2>0$ & $c_\Omega^2>0$ \\ \hline \xrowht{12pt}
    Odd-parity modes & ${\cal G}>0$, $G_{2,F}>0$ & ${\cal F}>0$ & ${\cal H}>0$ \\ \hline\xrowht{12pt}
    Even-parity modes & ${\cal K}>0$ & $c_{r2,{\rm even}}^2>0$ & $B_1^2\ge B_2>0$, $B_1<0$, $G_{2,F}+2FG_{2,FF}>0$ \\ \hline
\end{tabular}\label{table}
\end{table}

In \hyperref[table]{Table}, we summarize all the linear stability conditions 
in both odd- and even-parity sectors.
The radial propagation speeds of vector field perturbations 
$\delta A$ (odd-parity) and $V$ (even-parity) are both luminal 
($c_{r2,{\rm odd}}^2=c_{r3,{\rm even}}^2=1$).
In the gravitational sector, the radial propagation speeds 
of $\chi$ (odd-parity) and $\psi$ (even-parity) are 
equivalent to each other 
($c_{r1,{\rm odd}}^2=c_{r1,{\rm even}}^2={\cal G}/{\cal F}$).
We note that, on the FLRW background, the nonlinear term of $F$ 
in $G_2$ does not affect the perturbation dynamics by reflecting the fact 
that the quantity $F$ vanishes in $U(1)$ gauge-invariant 
theories \cite{Kase:2018nwt}. 
In contrast, the quantity $F$ does not vanish on the static and 
spherically symmetric background, and nonlinear terms of $F$ 
affect linear stability conditions in the odd- and even-parity sectors.

\subsubsection{$l=0$}

We proceed to the analysis of the monopole perturbation $l=0$, i.e., 
$L=0$. In this case, the perturbations $h_0$, $h_1$, and $G$ vanish 
identically from the second-order action of 
even-parity perturbations~\cite{Kobayashi:2014wsa,Kase:2021mix}. 
While one can choose the gauge different from Eq.~(\ref{gauge1}) 
to eliminate perturbations other than $h_0$ and $G$, we avoid doing 
so since the gauge DOFs are not  completely fixed in such a case. 
For $l=0$, the total action (\ref{acteven}) reduces to 
\ba
{\cal S}_{\rm even} 
&=& \sum_l\int {\rm d}t\, {\rm d}r 
\Bigg\{
v_1\left[2V\left(\delta A_0'-\dot{\delta A}_1+
\frac{v_3H_2+v_4\delta\phi'+v_5\delta\phi}{2v_1}\right)
-V^2\right]
-\frac{\left(v_3H_2+v_4\delta\phi'+v_5\delta\phi\right)^2}{4v_1}
\notag\\
&&
+\left(\Phi'+A_0'v_1V\right)H_0
-\frac{2}{f}\dot{\Phi}H_1
+c_1 \dot{\delta \phi} \dot{H}_2
+\left( c_2 \delta \phi'+ c_3 \delta \phi\right)H_2
+c_6H_2^2
+e_1 \dot{\delta \phi}^2+e_2 \delta \phi'^2
+e_3 \delta \phi^2
\Bigg\}
\,,
\label{acteven0}
\ea
where we introduced the combination 
\be
\Phi \equiv a_1\delta\phi'+\left(a_2-a_1'-\frac12A_0'v_4\right)\delta\phi+a_3H_2\,, 
\label{defPhi}
\ee
and used the relations among coefficients given in Appendix~\ref{AppA}. 
The quadratic terms of $H_1$, $\delta A_0$, and $\delta A_1$ present 
in the original Lagrangians (\ref{Lu}) and (\ref{LdA2}) disappear in Eq.~(\ref{acteven0}). 
This means that, in addition to $H_0$, each perturbation, $H_1$, $\delta A_0$, 
$\delta A_1$, plays a role of the Lagrange multiplier for $l=0$, and  
their Euler-Lagrange equations put constraints on other variables. 
Indeed, varying (\ref{acteven0}) with respect to $\delta A_0$ 
and $\delta A_1$ leads to 
\be
\left(v_1V\right)'=0\,,\qquad
\left(v_1V\right)\dot{}=0\,,
\ee
respectively. They are integrated to give 
\be
v_1V={\cal C}_1\,, 
\label{solV}
\ee
where ${\cal C}_1$ is a constant. 
This shows that $V$ depends only on $r$ and hence 
it is nondynamical for $l=0$. 
On the other hand, varying the action (\ref{acteven0}) 
with respect to $H_0$ and $H_1$ gives 
\be
\Phi'+A_0'v_1V=0\,,\qquad
\dot{\Phi}=0\,,
\ee
respectively. Integrating these two equations with the use 
of Eq.~(\ref{solV}), we obtain 
\be
\Phi={\cal C}_2-{\cal C}_1A_0\,,
\label{solPhi}
\ee
where ${\cal C}_2$ is an integration constant. 
Combining Eq.~(\ref{defPhi}) with 
Eq.~(\ref{solPhi}), 
we find that the perturbation $H_2$ can be expressed by using 
other variables as 
\be
H_2=-\frac{1}{a_3}\left[a_1\delta\phi'+\left(a_2-a_1'-\frac12A_0'v_4\right)\delta\phi
-({\cal C}_2-{\cal C}_1A_0)\right]\,.
\label{solH2}
\ee
In the present case, the perturbation (\ref{defpsi}) 
reduces to 
$\psi=H_2+a_1\delta \phi'/a_3$ and hence Eq.~(\ref{solH2}) gives 
a constraint on $\psi$. This means that, for $l=0$, the gravitational 
perturbation $\psi$ is not a propagating DOF.

Substituting Eqs.~(\ref{solV}), (\ref{solPhi}), and (\ref{solH2}) 
together with $\dot{H}_2$ 
into Eq.~(\ref{acteven0}), the resulting second-order 
action consists of $\delta\phi$ and its derivatives.
Since the integration constants ${\cal C}_1$ and ${\cal C}_2$ 
are irrelevant to the dynamics of perturbations, 
we simply set ${\cal C}_1=0={\cal C}_2$ in the following discussion.
After the integration by parts, we obtain the second-order 
action in the form 
\be
{\cal S}_{\rm even} 
=\int {\rm d}t\, {\rm d}r 
\left(
K_0\dot{\delta\phi}^2+G_0\delta\phi'^2
+M_0\delta\phi^2
\right)
\,,
\ee
where $K_0$, $G_0$, and $M_0$ are composed of 
the background quantities 
with the superscript representing $l=0$. 
This reduced action shows that the monopole perturbation 
possesses only one propagating DOF $\delta\phi$. 
The ghost is absent under the condition
\be
K_0=\frac{2{\cal P}_1-{\cal F}}{\sqrt{fh}\phi'^2}>0\,, 
\ee
which is equivalent to the no-ghost condition (\ref{nogoeven}) 
derived for $l\geq2$. 
The squared propagation speed $c_{r,{\rm even}}^2=-G_0/(fhK_0)$ 
also coincides with Eq.~(\ref{cr2even}) obtained for $l\geq2$. 
Consequently, the monopole perturbation $l=0$ does not 
give rise to additional stability conditions to those 
given in \hyperref[table]{Table}.

\subsubsection{$l=1$}

For the dipole mode $l=1$, the perturbations $K$ and $G$ appear in the second-order action only in 
the form 
$G-K$~\cite{Kobayashi:2014wsa,Kase:2021mix}. 
If we impose the gauge conditions $h_0=0$ and $K=G$, 
there is a residual gauge DOF associated with the transformation 
scalar ${\cal R}$. This can be fixed by choosing the gauge 
$\delta \phi=0$.
Namely, for $l=1$, we choose the gauge conditions
\be
h_0=0\,,\qquad K=G\,,\qquad
\delta\phi=0\,.
\ee
Eliminating nondynamical variables from the action (\ref{acteven}) 
with the approach analogous to 
the case $l \geq 2$, 
the second-order action can be expressed in the form (\ref{evenact}) 
with two dynamical perturbations 
\be
\vec{\mathcal{X}}=\left( 
\begin{array}{c}
\psi\\
V
\end{array}
\right) \,.
\ee
The ghosts are absent under the conditions 
\be
K_{22}>0\,,\qquad
K_{11} K_{22} -K_{12}^2>0\,,
\ee
which are equivalent to Eqs.~(\ref{K33}) and (\ref{detK1}), respectively, 
with the substitution of $L=2$.
The propagation speeds $c_r$ 
along the radial direction obey the same dispersion relation
as Eq.~(\ref{dispr}).
On using the properties $G_{12}=-fhK_{12}$ and $G_{22}=-fhK_{22}$, 
it follows that the squared propagation speeds of $\psi$ 
and $V$ 
are identical to Eqs.~(\ref{cr2even}) and 
(\ref{cr3even}), respectively. 
Thus, the dipole perturbation possesses two propagating DOFs 
arising from the scalar and vector field perturbations. 
We do not have additional conditions to those given 
in \hyperref[table]{Table}.

\section{Application to concrete theories with hairy BHs}
\label{appsec}

Theories with the action (\ref{action}) can accommodate 
a wide variety of hairy BH solutions known in the literature.
In this section, we apply the linear stability conditions derived 
in Secs.~\ref{oddsec} and \ref{evensec} to concrete BHs 
present in the framework of Maxwell-Horndeski theories. 
We will focus on the case 
$l \geq 2$, in which case five 
dynamical DOFs are present.

\subsection{Nonminimally coupled k-essence with a gauge field}

We begin with a subclass of Maxwell-Horndeski given by the action   
\be
{\cal S} =
\int {\rm d}^4 x \sqrt{-g}\,
\left[ G_2(\phi,X,F)+G_{4}(\phi) R \right]\,,
\label{model1}
\ee
where the nonminimal coupling $G_4$ is a function of $\phi$ only. 
The analysis of BH perturbations in this case was also 
addressed in Ref.~\cite{Gannouji:2021oqz}, 
but the angular stability conditions of even-parity perturbations were missing. 
In the following, we will investigate all the linear 
stability conditions.

The stability of odd-parity perturbations requires that 
\be
{\cal G}={\cal F}={\cal H}=2G_4>0\,,\qquad 
G_{2,F}>0\,.
\label{oddmo1}
\ee
Hence all the propagation speeds of $\chi$ and $\delta A$ 
are luminal in both radial and angular directions.
In the even-parity sector, the quantity (\ref{nogoeven}) yields
\be
{\cal K}=\frac{2 r^2 \phi'^2 G_4 
(G_{2,X}G_4+3G_{4,\phi}^2)}
{(2G_4+r \phi' G_{4,\phi})^2}\,.
\ee
If the BH has a scalar hair, the field derivative $\phi'$ 
is nonvanishing. Under the first inequality (\ref{oddmo1}), 
the no-ghost condition ${\cal K}>0$ translates to 
\be
G_{2,X}G_4+3G_{4,\phi}^2>0\,.
\label{evenmo1}
\ee
For a minimally coupled scalar field 
($G_4={\rm constant}>0$), 
this condition 
reduces to $G_{2,X}>0$. 
The radial propagation speeds of both $\psi$ and $V$ 
are luminal ($c_{r1,{\rm even}}^2=c_{r3,{\rm even}}^2=1$). 
To ensure the Laplacian stability of $\delta \phi$ along the 
radial direction, we require that  
\be
c_{r2,{\rm even}}^2=1+\frac{2G_4 X 
[G_{2,XX}(G_{2,F}+2FG_{2,FF})-2 F G_{2,FX}^2]}
{(G_{2,X}G_4+3G_{4,\phi}^2)(G_{2,F}+2FG_{2,FF})}>0\,,
\label{cr2mo1}
\ee
which coincides with Eq.~(5.26) of Ref.~\cite{Gannouji:2021oqz} 
(one needs to replace 
$G_{2} \to f_2$ and $G_4 \to f_1/2$ for the notation used in \cite{Gannouji:2021oqz}).

Along the angular direction, the quantities (\ref{B1def}) and 
(\ref{B2def}) reduce, respectively, to 
\be
B_1=-1\,,\qquad B_2=1\,,
\ee
and hence 
\be
c_{\Omega+,{\rm even}}^2=1\,,\qquad
c_{\Omega-,{\rm even}}^2=1\,.
\ee
Thus, there are no angular Laplacian instabilities
for the perturbations $\psi$ and $\delta \phi$. 
The angular stability of vector field perturbation $V$ 
requires that 
\be
c_{\Omega3,{\rm even}}^2=
\frac{G_{2,F}}{G_{2,F}+2F G_{2,FF}}>0\,.
\label{cO3b}
\ee
Under the no-ghost condition $G_{2,F}>0$, 
the inequality (\ref{cO3b}) is satisfied 
for $G_{2,F}+2F G_{2,FF}>0$. 

In the following, we will study the stability of hairy BHs in two 
subclasses of theories given by the action (\ref{model1}).

\subsubsection{Einstein-Maxwell-dilaton theory}
\label{GMsec}

In bosonic heterotic string 
theory, the gauge field is coupled 
to a dilaton field $\phi$ with the 
Lagrangian $4e^{-2\phi}F$. 
In the Einstein frame, the 
corresponding effective four-dimensional action 
is given by 
\be
{\cal S} =
\int {\rm d}^4 x \sqrt{-g}\,
\left( R+4X+4e^{-2\phi}F \right)\,,
\label{GM}
\ee
where the unit $\Mpl^2/2=1$ is used, 
with $\Mpl$ being the reduced Planck mass. 
In this theory, there is an exact BH solution advocated by GM 
and GHS \cite{Gibbons:1987ps,Garfinkle:1990qj}. 
GHS derived the hairy BH solution for a static and 
spherically symmetric metric where $r^2$ in front of 
$\rd \Omega^2$ in Eq.~(\ref{metric}) is modified to 
a general function $\zeta^2(r)$. 
In Appendix \ref{AppB}, we revisit the derivation 
of this exact solution.
In terms of the coordinate (\ref{metric}), the GM-GHS  
solution is expressed as 
\ba
& &
f=1-\frac{2M}{r^2} \left( \sqrt{r^2+r_q^2}-r_q \right)\,,\qquad 
h=\left( 1+\frac{r_q^2}{r^2} \right)f\,,\nonumber \\
& &
\phi=\phi_0+\frac{1}{2} \ln \left( 
\frac{\sqrt{r^2+r_q^2}-r_q}{\sqrt{r^2+r_q^2}+r_q} \right)
\,,\qquad 
A_0'=\frac{qr e^{2\phi_0}}{\left( \sqrt{r^2+r_q^2}+r_q \right)^2
\sqrt{r^2+r_q^2}}\,,
\label{GMso}
\ea
where $M$ is a constant, $\phi_0$ is an asymptotic value 
of $\phi$ at spatial infinity, and 
\be
r_q \equiv \frac{q^2 e^{2\phi_0}}{2M}\,.
\label{rq}
\ee
Here, $q$ is a constant corresponding to an electric charge. 
The radial derivative of $\phi$ is given by $\phi'(r)=r_q/[r\sqrt{r^2+r_q^2}]$, 
which behaves as $\phi'(r) \simeq q^2 e^{2\phi_0}/(2M r^2)$ at 
spatial infinity. The scalar field acquires a secondary hair $q$ 
through a dilatonic coupling with the gauge field. 
In the limit that $q \to 0$, the solution (\ref{GMso}) reduces to the 
Schwarzschild metric $f=h=1-2M/r$ with $\phi=\phi_0$ and 
$A_0'=0$. For $q \neq 0$, there is a single event 
horizon  \cite{Garfinkle:1990qj} located at 
\be
r_H=2\sqrt{M(M-r_q)}\,, 
\ee
whose existence requires that $r_q<M$. 
{}From Eq.~(\ref{GMso}), both $\phi'$ and $A_0'$ are finite at $r=r_H$.

The action (\ref{GM}) corresponds to the coupling functions
\be
G_2=4X+4e^{-2\phi}F\,,\qquad G_4=1\,,
\ee
in Eq.~(\ref{model1}).
In this case we have ${\cal G}={\cal F}={\cal H}=2$, 
$G_{2,F}=4e^{-2\phi}$, and $G_{2,X}G_4+3G_{4,\phi}^2=4$, 
so the conditions (\ref{oddmo1}) and (\ref{evenmo1}) 
are automatically satisfied. 
The two squared propagation speeds (\ref{cr2mo1}) 
and (\ref{cO3b}) reduce, respectively, to 
\be
c_{r2,{\rm even}}^2=1\,,\qquad c_{\Omega3,{\rm even}}^2=1\,.
\ee
Thus, all the linear stability conditions are consistently satisfied 
for the GM-GHS BH solution.

We can consider more general theories in which the dilatonic 
coupling $e^{-2\phi}$ is extended to an arbitrary function $\xi$ of $\phi$, i.e., 
$G_2=4X+4\xi(\phi)F$ and 
$G_4=1$. This includes the case of spontaneous scalarized 
BHs which can be realized for 
even-power law functions of 
$\xi(\phi)$ \cite{Herdeiro:2018wub,Fernandes:2019rez,Myung:2018jvi,Blazquez-Salcedo:2020nhs,Konoplya:2019goy}.
In such Einstein-Maxwell-scalar theories, the difference of stability conditions from the dilatonic case appears only for the 
quantity $G_{2,F}=4\xi(\phi)$. 
So long as $\xi(\phi)>0$,  
hairy BH solutions are consistent with all the linear stability conditions. 

\subsubsection{Einstein-Born-Infeld-dilaton gravity}

A BI-type action can arise as a low energy 
effective action describing the dynamics of vector fields 
in open string theory or on D-branes \cite{Tseytlin:1986ti,Fradkin:1985qd,Abouelsaood:1986gd,Leigh:1989jq,
Cederwall:1996uu,Gibbons:1995cv,BeltranJimenez:2017doy}.
The Lagrangian of such a nonlinear BI theory is given by 
${\cal L}_{\rm BI}(F)=(4/b^2)[1-\sqrt{1-2b^2 F}]$, where 
$b$ is a coupling constant. 
The nonlinear BI vector field can be coupled to the dilaton field $\phi$.
The action of Einstein-BI-dilaton theory is given by 
\be
{\cal S} =
\int {\rm d}^4 x \sqrt{-g}\,
\left[ R+\eta X+\frac{4}{b^2 \mu(\phi)}
\left( 1-\sqrt{1-2b^2 \mu^2(\phi) F} \right) \right]\,,
\label{BIa}
\ee
where $\eta$ is a constant, and $\mu(\phi)=e^{-2\phi}$.
This theory corresponds to the coupling functions
\be
G_2=\eta X+\frac{4}{b^2 \mu(\phi)}
\left( 1-\sqrt{1-2b^2 \mu^2(\phi) F} \right)\,,\qquad 
G_4=1\,,
\label{G24}
\ee
where $1-2b^2 \mu^2(\phi) F>0$ 
for theoretical consistency. 
In the limit that $b \to 0$ we have $G_2 \to \eta X+4\mu(\phi)F$, 
so it recovers the theory discussed in Sec.\,\ref{GMsec}.
In the regime of small values of $b$, there should be 
hairy BHs similar to the exact solution (\ref{GMso}). 
Indeed, the existence of regular BH solutions was shown 
for arbitrary couplings $b$ \cite{Clement:2000ue,Tamaki:2001vv,Yazadjiev:2005za,Sheykhi:2006ji,Stefanov:2007qw}.

Let us now discuss the linear stability of BHs in theories 
given by the coupling functions (\ref{G24}).
First of all, we have ${\cal G}={\cal F}={\cal H}=2>0$.
Since $G_{2,F}=4\mu(\phi)/\sqrt{1-2b^2 \mu^2(\phi) F}$, 
the dilatonic coupling $\mu(\phi)=e^{-2\phi}$ satisfies the 
condition $G_{2,F}>0$. 
The no-ghost condition (\ref{evenmo1}) yields
$G_{2,X}G_4+3G_{4,\phi}^2=\eta>0$.
The radial propagation speed squared (\ref{cr2mo1})
reduces to the luminal value $c_{r2,{\rm even}}^2=1$. 
On the other hand, the angular propagation speed 
squared (\ref{cO3b}) yields
\be
c_{\Omega3,{\rm even}}^2=1-b^2 \mu^2(\phi) \frac{h}{f}A_0'^2\,.
\label{cOBI}
\ee
{}From Eq.~(\ref{A0dso}), we have 
\be
A_0'=\frac{q \sqrt{f}}{\sqrt{h(r^4+b^2 q^2)} \mu(\phi)}\,,
\label{A0BI}
\ee
where $q=q_0/4$. Then, Eq.~(\ref{cOBI}) reduces to 
\be
c_{\Omega3,{\rm even}}^2=\frac{r^4}
{r^4+b^2 q^2}\,,
\ee
which is positive. Moreover, this vector field propagation speed is 
in the subluminal range $0<c_{\Omega3,{\rm even}}^2<1$. 
In summary, provided that $\eta>0$, all the linear stability conditions are consistently satisfied.

Finally, there is a specific case in which the scalar field $\phi$ 
is absent, i.e., 
\be
\eta=0\,,\qquad \mu(\phi)=1\,,
\ee
in the action (\ref{BIa}). In this theory, there is an exact
BH solution given by \cite{Fernando:2003tz,Cai:2004eh,Dey:2004yt}
\be
f=h=1-\frac{2M}{r}+\frac{2 r^2}{3b^2}
-\frac{2}{r} \int^r \frac{\sqrt{\tilde{r}^4+b^2 q^2}}{b^2} 
{\rm d}\tilde{r}\,,
\label{BIso}
\ee
where $A_0'$ is given by Eq.~(\ref{A0BI}) with $\mu(\phi)=1$.
The absence of the scalar field means that we do not have the no-ghost 
condition ${\cal K}>0$ associated with the perturbation $\delta \phi$.
Since the other stability conditions are the same as those derived 
for the action (\ref{BIa}) with the replacement $\mu(\phi) \to 1$, 
there are neither ghost nor Laplacian instabilities. 
We note that $c_{\Omega3,{\rm even}}$ is again subluminal.

\subsection{Einstein-Maxwell-Dilaton-Gauss-Bonnet Theory}

In low energy effective heterotic string theory, the dilaton field $\phi$ is not only coupled to the electromagnetic field strength 
$F$ but also to a GB curvature invariant 
$R_{\rm GB}^2=R^2-4R_{\mu \nu}R^{\mu \nu}
+R_{\mu \nu \rho \lambda}R^{\mu \nu \rho \lambda}$ 
of the form $e^{-2\phi} R_{\rm GB}^2$, 
where $R_{\mu \nu}$ is the Ricci tensor 
and $R_{\mu \nu \rho \lambda}$ is the Riemann tensor. 
At leading order in the $\alpha'$ expansion, the low energy 
effective action of heterotic strings in the bosonic sector 
is given by \cite{Gross:1986iv,Fradkin:1984pq,Zwiebach:1985uq}
\be
{\cal S} =
\int {\rm d}^4 x \sqrt{-g}\,
\left[ R+4X+\alpha \xi(\phi) \left( 4F+
R_{\rm GB}^2\right) \right]\,,
\label{GM2}
\ee
where $\alpha=\alpha'/8$ is a coupling constant, and 
\be
\xi(\phi)=e^{-2\phi}\,.
\ee
It is known that hairy BH solutions are present in this 
theory \cite{Wiltshire:1985us,Mignemi:1992nt} (see also Refs.~\cite{Doneva:2018rou,Hunter:2020wkd} for 
recent related works).
The theory given by the action (\ref{GM2}) belongs to 
a subclass of Horndeski theories with the coupling functions \cite{KYY}
\ba
& &
G_2=4X+4\alpha \xi(\phi) F+8\alpha \xi_{,\phi \phi \phi \phi}(\phi)
X^{2}(3-\ln{|X}|)\,,\qquad 
G_3=4\alpha \xi_{,\phi \phi \phi}(\phi) X(7-3\ln{|X|})\,,
\nonumber \\
& &
G_4=1+4\alpha \xi_{,\phi \phi}(\phi) X(2-\ln{|X|})\,,
\qquad G_5=-4\alpha \xi_{,\phi}(\phi) \ln{|X|}\,.
\ea
In the limit that $\alpha \to 0$, the BH solution should reduce 
to the no-hair Schwarzschild metric.
For arbitrary couplings $\alpha$ it is difficult to derive an exact BH 
solution, but we can obtain solutions for small $\alpha$ 
by using the expansions
\ba
& &
f(r)=
\left(1-\frac{2M}{r}\right)
\biggl[
1+\sum_{j\ge 1}
\hat{f}_j(r) \alpha^j \biggr]\,,
\qquad
h(r)
=
\left(1-\frac{2M}{r}\right)
\biggl[
1+\sum_{j\ge 1}
\hat{h}_j(r)\alpha^j
\biggr]\,,\nonumber \\
& &
\phi(r)= \phi_0+
\sum_{j\ge 1}
\hat{\phi}_j(r) \alpha^j\,,
\label{sch_pert}
\ea
where $M$ and $\phi_0$ are constants, 
$\hat{f}_j(r)$, $\hat{h}_j(r)$, and $\hat{\phi}_j(r)$
are functions of $r$. 
The temporal vector component obeys Eq.~(\ref{A0dso}), i.e., 
\be
A_0'=e^{2\phi}\sqrt{\frac{f}{h}}\frac{q}{r^2}\,,
\label{A0GB}
\ee
where $q$ is a constant.
We substitute Eqs.~(\ref{sch_pert}), (\ref{A0GB}), and their $r$ derivatives 
into the background Eqs.~(\ref{back1})-(\ref{back3}) and 
(\ref{Ephi}). Then, we derive the solutions to 
$\hat{f}_j(r)$, $\hat{h}_j(r)$, and $\hat{\phi}_j(r)$ at each order in $\alpha$.

At first order in $\alpha$, the solutions regular on the 
horizon ($r=2M$) are given by 
\be
\hat{f}_1(r)=-\frac{q^2 e^{2\phi_0}}{2M^2 \hat{r}}\,,\qquad 
\hat{h}_1(r)=-\frac{q^2 e^{2\phi_0}}{2M^2 \hat{r}}\,,\qquad 
\hat{\phi}_1(r)=\tilde{\phi}_1
-\frac{3e^{2\phi_0}q^2 \hat{r}^2
+2e^{-2\phi_0}(3\hat{r}^2+3\hat{r}+4)}{6M^2 \hat{r}^3}\,,
\ee
where $\hat{r} \equiv r/M$, and $\tilde{\phi}_1$ is a constant.
For $q \to 0$, the vector field derivative (\ref{A0GB}) is vanishing 
and hence this corresponds to the limit in which only the 
dilaton-GB coupling $\alpha \xi(\phi) R_{\rm GB}^2$ is present. 
In this limit we have $\hat{f}_1(r)=0=\hat{h}_1(r)$, so the GB term 
does not contribute to the 
metric components at this order. 
This property is consistent with the findings in 
Refs.~\cite{Minamitsuji:2022mlv,Minamitsuji:2022vbi}.  
We note that $\hat{\phi}_1(r)$ is affected by both the GB term 
and the vector field.

At second order in $\alpha$, we obtain the following regular solutions
\ba
\hspace{-0.5cm}
\hat{f}_2(r) &=&
[3200+832 \hr+112( 5q^2 e^{4\phi_0}-1) \hr^2
-8\{ 137+5q^2 e^{4\phi_0} (6\tilde{\phi}_1 M^2 \hr^2 
e^{2\phi_0}+5) \} \hr^3\nonumber \\
& &
+6(5q^4 e^{8\phi_0}-10q^2 e^{4\phi_0}-98) \hr^4
+3(5q^4 e^{8\phi_0}-10q^2 e^{4\phi_0}-98) \hr^5 
]/(240e^{4\phi_0} M^4 \hr^6)\,,\\
\hspace{-0.5cm}
\hat{h}_2(r) &=&
[14720+6976 \hr
+16(125 q^2 e^{4\phi_0}+203) \hr^2
+24( 5q^2 e^{4\phi_0}-19) \hr^3
 \nonumber \\
& &
+6(15 q^4 e^{8\phi_0}+30q^2 e^{4\phi_0}-58) \hr^4
+3 \{ 5q^4 e^{8\phi_0}-10q^2 e^{4\phi_0}(8\tilde{\phi}_1 
M^2 e^{2\phi_0}+1)-98 \}\hr^5]/(240e^{4\phi_0} M^4 \hr^6)\,,\\
\hspace{-0.5cm}
\hat{\phi}_2(r) &=& \tilde{\phi}_2
+\tilde{\phi}_1\frac{8+6\hr+3(2-q^2e^{4\phi_0})\hr^2}
{3e^{2\phi_0}M^2\hr^3} \nonumber \\
& &
-[1600+2688 \hr+60(73-10 q^2 e^{4\phi_0}) \hr^2 
+40(73+15 q^2 e^{4\phi_0}) \hr^3+2190 \hr^4 \nonumber \\
& &+15(146-30q^2e^{4\phi_0}-15q^4 e^{8\phi_0}) \hr^5]
/(1800e^{4\phi_0}M^4 \hr^6)\,,
\ea
where $\tilde{\phi}_2$ is a constant. 
At this order, both the GB term and the vector field contribute to the
metric components. 
At spatial infinity, all of $\hat{f}_1(r)$, $\hat{h}_1(r)$, $\hat{f}_2(r)$, 
$\hat{f}_2(r)$ approach 0 with $\hat{\phi}_1(r) \to \tilde{\phi}_1$ 
and $\hat{\phi}_2(r) \to \tilde{\phi}_2$, so the above hairy BH solutions 
are asymptotically flat.
 
We derive the solutions (\ref{sch_pert}) expanded 
up to the sixth order in $\alpha$ and use them 
to compute quantities associated with the linear stability of BHs. 
Since $G_{2,F}=4\alpha e^{-2\phi}$, we require the condition 
\be
\alpha>0\,,
\ee
to avoid ghost instabilities of vector field perturbations.  
The angular propagation speed 
of $V$ is luminal,
$c_{\Omega3,{\rm even}}^2=1$. 
In the odd-parity sector, we have 
\ba
{\cal G} &=& 2+\frac{8[q^2 \hat{r}^2+2e^{-4\phi_0} 
(\hat{r}^2+2 \hat{r}+4)]}
{M^4 \hat{r}^6}\alpha^2+{\cal O} (\alpha^3)\,,\label{GGB} \\
{\cal F} &=& 2-\frac{8[q^2 \hat{r}^2(2\hat{r}-5)
+2e^{-4\phi_0} (2\hat{r}^3+\hat{r}^2+2 \hat{r}-36)]}
{M^4 \hat{r}^6}\alpha^2+{\cal O} (\alpha^3)\,,\\
{\cal H} &=& 2+\frac{8(\hat{r}-2)[q^2 \hat{r}^2
+2e^{-4\phi_0}(\hat{r}^2+2\hat{r}+4)]}
{M^4 \hat{r}^6}\alpha^2+{\cal O} (\alpha^3)\,.
\label{HGB}
\ea
In the small coupling regime where terms of order $\alpha^2$ in Eqs.~(\ref{GGB})-(\ref{HGB}) are smaller than the order 1, 
all of ${\cal G}$, ${\cal F}$, and ${\cal H}$ are positive.

In the even-parity sector, the no-ghost parameter (\ref{nogoeven}) yields
\be
{\cal K}=\frac{[e^{2\phi_0}q^2 \hat{r}^2
+2e^{-2\phi_0}(\hat{r}^2+2\hat{r}+4) ]^2}
{2M^4 \hat{r}^6}\alpha^2+{\cal O}(\alpha^3)\,,
\ee
whose leading-order term is always positive. 
The radial squared propagation speed of $\delta \phi$ can be estimated as 
\be
c_{r2,{\rm even}}^2=
1-\frac{32(\hat{r}-2)
[q^4 \hat{r}^4+4q^2 \hat{r}^2
(\hat{r}^2+2\hat{r}+16)e^{-4\phi_0}
+4(\hat{r}^4+4\hat{r}^3+36 \hat{r}^2+88 \hat{r}+208)
e^{-8\phi_0}]}{M^8 \hat{r}^{12}}\alpha^4
+{\cal O}(\alpha^5)\,.
\label{cr6th}
\ee
To derive this result, we need to use the solutions (\ref{sch_pert}) expanded 
up to the order $j=6$. 
The solutions expanded up to 
$j=7$ give the same coefficient 
of $\alpha^4$ in $c_{r2,{\rm even}}^2$ 
as that appearing in 
Eq.~(\ref{cr6th}).
For small $\alpha$, $c_{r2,{\rm even}}^2$ 
is close to 1 and hence there is no Laplacian stability 
of $\delta \phi$ along the radial direction. 
The squared propagation speeds of $\psi$ and 
$\delta \phi$ along the angular direction are given by 
\be
c_{\Omega\pm,{\rm even}}^2=1 \pm \frac{24e^{-2\phi_0}}
{M^2 \hat{r}^3}
\alpha+{\cal O}(\alpha^2)\,, 
\ee
where terms of order $\alpha$ 
arises from the dilaton-GB 
coupling.
For small $\alpha$, both 
$c_{\Omega+,{\rm even}}^2$ and $c_{\Omega-,{\rm even}}^2$ 
are positive. We have thus shown that all the linear stability 
conditions are consistently satisfied for hairy BH solutions 
present for small couplings $\alpha$.

\subsection{4DEGB gravity}

The GB curvature invariant $R_{\rm GB}^2$ is a topological surface term, 
so the field equations of motion following from the action
${\cal S}=\int {\rm d}^4 x
\sqrt{-g}\,\hat{\alpha}_{\rm GB} R_{\rm GB}^2$ vanish 
in 4 dimensions \cite{Lovelock:1971yv}. 
In a $D$-dimensional spacetime ($D>4$), rescaling the GB coupling constant 
as $\hat{\alpha}_{\rm GB} \to \alpha_{\rm GB}/(D-4)$ allows a possibility 
for extracting contributions of the higher-dimensional GB term \cite{Glavan:2019inb}. 
One can perform a Kaluza-Klein reduction of the $D$-dimensional 
Einstein-GB gravity on a $(D-4)$-dimensional maximally symmetric space with 
a vanishing spatial curvature \cite{Lu:2020iav,Kobayashi:2020wqy}. 
The size of a maximally symmetric space
is characterized by a scalar field $\phi$. 
Taking the Maxwell field into account, the four-dimensional action 
derived from the Kaluza-Klein reduction of $D$-dimensional 
Einstein-GB theory belongs to a subclass of shift-symmetric 
Maxwell-Horndeski theories given by 
the coupling functions \cite{Fernandes:2022zrq}
\be
G_2=8 \alpha_{\rm GB} X^2+4F\,,\qquad 
G_3=8 \alpha_{\rm GB} X\,,\qquad 
G_4=1+4\alpha_{\rm GB}X\,,\qquad 
G_5=4 \alpha_{\rm GB} \ln |X|\,.
\ee

In this regularized 4DEGB gravity, it is known that there exists 
an exact hairy BH solution \cite{Fernandes:2020rpa}.
We first revisit the derivation of this exact BH solution 
and then study its linear stability. 
The background equations of motion for the line element  
(\ref{metric}) are expressed in the form 
\ba
& &
f'= -\frac{r^2 f(h-1)+\alpha_{\rm GB} 
f[h^2-2h (1-2j-2r \phi' j)
+1-4j+3j^2]+h r^4 A_0'^2}
{hr[r^2-2\alpha_{\rm GB} (h-1+j+r\phi'j)]}\,,
\label{feq}\\
& &
\frac{h'}{h}-\frac{f'}{f}=-\frac{4\alpha_{\rm GB} rj 
(\phi'^2-\phi'')}
{r^2-2\alpha_{\rm GB} (h-1+j+r \phi' j)}\,,
\label{heq} \\
& &
\sqrt{\frac{h}{f}}\alpha_{\rm GB} \left( f'+2\phi' f \right) j
={\cal C}\,,
\label{Jeq}\\
& &
A_0'=\sqrt{\frac{f}{h}}\frac{q}{r^2}\,,
\label{A0d4}
\ea
where ${\cal C}$ and $q$ are constants, and 
\be
j \equiv 1-h (1+r \phi')^2\,.
\ee
We search for asymptotically flat BH solutions with 
$f=f_0+f_1/r+f_2/r^2+\cdots$, 
$h=1+h_1/r+h_2/r^2+\cdots$, and 
$\phi=\phi_0+\phi_1/r+\phi_2/r^2+\cdots$ 
at spatial infinity, where $f_j$, $h_j$, and $\phi_j$ 
are constants. The left hand-side of 
Eq.~(\ref{Jeq}) approaches 0 as $r \to \infty$, 
so the constant ${\cal C}$ is fixed to be 0.
Since $f$, $h$, $f'$, and $\phi$ are finite outside the horizon, 
we require that $j=0$. Then, the scalar field solution with 
the behavior $\phi' \propto 1/r^2$ at large distances 
is given by 
\be
\phi'=\frac{1}{r} \left( \frac{1}{\sqrt{h}}-1 \right)\,,
\label{phi4}
\ee
which diverges on the horizon ($h=0$). 
We note that the field kinetic term 
$X=-(1-\sqrt{h})^2/(2r^2)$ is finite at $h=0$. 
We can integrate Eq.~(\ref{heq}) to give 
$h=\tilde{{\cal C}} f$, 
where $\tilde{{\cal C}}$ is a constant. 
Since $\tilde{{\cal C}}$ can be chosen 1 after the time 
reparametrization of $f$, it follows that $h=f$. 
Substituting $j=0$, $h=f$, and Eq.~(\ref{A0d4}) into 
Eq.~(\ref{feq}), we obtain
\be
f'=-\frac{(f-1)[r^2+\alpha_{\rm GB} (f-1)]+q^2}
{r^3-2\alpha_{\rm GB} r (f-1)}\,.
\label{df4}
\ee
The integrated solution to this equation, which 
is consistent with the boundary condition at 
spatial infinity, is given by 
\be
f=h=1+\frac{r^2}{2\alpha_{\rm GB}} \left[ 1
-\sqrt{1+4\alpha_{\rm GB} \left( \frac{2M}{r^3}
-\frac{q^2}{r^4} \right)} \right]\,,
\label{fh4}
\ee
where $M$ is a constant. 
The vector field solution (\ref{A0d4}) reduces to $A_0'=q/r^2$.
The horizons are located at 
\be
r_{\pm}=M \pm \sqrt{M^2-q^2-\alpha_{\rm GB}}\,.
\ee
The existence of horizons requires the condition 
$q^2+\alpha_{\rm GB} \le M^2$.

On using Eqs.~(\ref{A0d4}), (\ref{phi4}), (\ref{df4}) with $f=h$, 
it follows that 
\be
{\cal K}=0\,,
\label{K0}
\ee
at any distance $r$. Since the dynamical scalar field $\phi$ 
is present as a radion mode in the extra dimension, the 
vanishing kinetic term ${\cal K}$ means a strong coupling 
problem associated with the perturbation $\delta \phi$. 
The same strong coupling problem is also present for hairy BH 
solutions with $q=0$ \cite{Tsujikawa:2022lww}. 
The denominator of Eq.~(\ref{cr2even}) is proportional to 
${\cal K}=2{\cal P}_1-{\cal F}$ and hence $c_{r2,{\rm even}}^2$ 
is divergent for arbitrary $r$. 
We also note that both $B_1$ and $B_2$ contain 
${\cal K}$ in their denominators, so this generally leads to 
the divergences of $c_{\Omega\pm,{\rm even}}^2$ as well.

We can compute the ratio $c_{r2,{\rm even}}^2/B_2$ at large distances 
by using the following asymptotic solution of Eq.~(\ref{fh4}),
\be
f=h=1-\frac{2M}{r}+\frac{q^2}{r^2}+\frac{4\alpha_{\rm GB} M^2}{r^4}
-\frac{4\alpha_{\rm GB} q^2 M}{r^5}+{\cal O}(r^{-6})\,.
\ee
Then, we obtain the following asymptotic behavior
\be
\frac{c_{r2,{\rm even}}^2}{B_2}=-2-\frac{3(3M^2+q^2)}{2Mr}
+{\cal O}(r^{-2})\,.
\label{crB}
\ee
This means that both $c_{r2,{\rm even}}^2$ and $B_2$ cannot be 
simulataneously positive at large distances, 
so either of the linear stability conditions 
$c_{r2,{\rm even}}^2>0$ or $B_2>0$ is violated.
The same instability was also found for hairy BH solutions with 
$q=0$ \cite{Tsujikawa:2022lww}.

In the vicinity of the outer horizon $r_+=M+\sqrt{M^2-q^2-\alpha_{\rm GB}}$, 
the metric components can be expanded as
\ba
& &
f=h=\frac{r_+^2-q^2-\alpha_{\rm GB}}{r_+ (r_+^2+2\alpha_{\rm GB})}(r-r_+)
-\frac{r_+^6-(2q^2+3\alpha_{\rm GB})r_+^4-2\alpha_{\rm GB}
(q^2+3\alpha_{\rm GB})r_+^2
-\alpha_{\rm GB}
(q^2+\alpha_{\rm GB})^2}{r_+^2(r_+^2+2\alpha_{\rm GB})^3}(r-r_+)^2 
\nonumber \\
& &\qquad \qquad+{\cal O}(r-r_+)^3\,.
\label{fexpan}
\ea
On using this expanded solution with Eqs.~(\ref{A0d4}) and (\ref{phi4}), 
the product ${\cal F}{\cal K}B_2$ yields
\be
{\cal F}{\cal K}B_2=
-\frac{16\alpha_{\rm GB}^2 [r_+^4+(r_+^2+q^2)\alpha_{\rm GB}
+\alpha_{\rm GB}^2]^2}
{r_+^2 (r_+^2+2\alpha_{\rm GB})^4}(r-r_+)^{-2}
+{\cal O} \left( (r-r_+)^{-3/2} \right)\,.
\ee
Hence the leading-order term of ${\cal F}{\cal K}B_2$ is negative 
around $r=r_+$. 
This means that either ${\cal F}$, ${\cal K}$, or $B_2$ must 
be negative, so the hairy BH is unstable in the vicinity of 
the outer horizon. 
For $q=0$, the leading-order term of ${\cal F}{\cal K}B_2$ 
coincides with the one derived in Ref.~\cite{Tsujikawa:2022lww}.

For the specific case $r_+^2=q^2+\alpha_{\rm GB}$, there is a 
single horizon located at $r_+=M$. Since the first term 
on the right hand-side of Eq.~(\ref{fexpan}) vanishes in such 
a case, the leading-order contribution to ${\cal F}{\cal K}B_2$ is not 
necessarily negative. However the properties (\ref{K0}) 
and (\ref{crB}) still hold, so the problems of strong 
coupling as well as large-distance Laplacian instability 
are unavoidable for the hairy BH present in regularized 
4DEGB gravity.

\section{Conclusions}
\label{consec}

In Maxwell-Horndeski theories given by the action (\ref{action}), we derived BH linear stability conditions on the static and spherically symmetric 
background (\ref{metric}). 
We incorporated a $U(1)$ gauge-invariant vector field $A_{\mu}$ coupled to a scalar field $\phi$ with the Lagrangian $G_2(\phi, X, F)$, where 
$F=-F_{\mu \nu}F^{\mu \nu}/4$ is the gauge field strength. 
Due to the gauge invariance, the vector field equation has an integrated solution of the form (\ref{A0dso}), where $q_0$ corresponds to an electric charge. 
As we observe in Eqs.~(\ref{back1}) and (\ref{back2}), the temporal 
component of $A_{\mu}$ modifies the background equations in the 
gravitational sector. The scalar field Eq.~(\ref{Ephi}) can be also 
generally affected by the coupling with $A_{\mu}$ through the existence of 
$G_2$-dependent terms 
such as $G_{2,\phi}$. 

In Sec.~\ref{oddsec}, we first showed that the second-order Lagrangian of odd-parity perturbations is expressed as Eq.~(\ref{Lodd}). 
Introducing an auxiliary field $\chi$ defined by Eq.~(\ref{chi}), 
the Lagrangian for the multipoles $l \geq 2$ can be expressed 
in terms of the two dynamical fields $\chi$ and $\delta A$.
They correspond to the perturbations arising from the gravitational 
and vector field sectors, respectively. 
We found that the propagation of $\chi$ 
is analogous to the case of Horndeski theories without the 
vector field \cite{Kobayashi:2012kh}. 
In the limit $l \gg 1$, 
the propagation speeds of $\delta A$ are luminal in both radial 
and angular directions.
In the odd-parity sector, there are neither ghost nor Laplacian 
instabilities under the conditions ${\cal G}>0$, $G_{2,F}>0$, 
${\cal F}>0$, and ${\cal H}>0$. 
For $l=1$, $\delta A$ is the only propagating DOF,  
whose stability does not require additional conditions.

In Sec.~\ref{evensec}, we obtained the second-order action of 
even-parity perturbations in the form (\ref{acteven}) with 
the sum of (\ref{Lu}) and (\ref{LdA}). 
The auxiliary field $V$ introduced in Eq.~(\ref{Vexp}) plays 
a role of the dynamical vector field perturbation in the 
even-parity sector. There are also two dynamical perturbations 
$\psi$ and $\delta \phi$ arising from the gravitational and 
scalar field sectors, respectively. 
For $l \geq 2$, we showed that the second-order action 
can be expressed in the form (\ref{evenact}) with 
$\vec{\mathcal{X}}={}^t (\psi, \delta \phi, V)$.
Under the stability conditions ${\cal F}>0$ and $G_{2,F}>0$ 
for odd-parity perturbations, 
the ghosts in the even-parity sector
are absent under the condition ${\cal K}=2{\cal P}_1-{\cal F}>0$, where ${\cal P}_1$ 
is defined in Eq.~(\ref{defP1P2}). 

The squared propagation speeds of even-parity perturbations 
$\psi$, $\delta \phi$, $V$ along the radial direction are given, 
respectively, by Eqs.~(\ref{cr1even}), (\ref{cr2even}), and (\ref{cr3even}).
While the expression of $c_{r1,{\rm even}}$ coincides with that derived 
in Ref.~\cite{Kobayashi:2014wsa,Kase:2021mix}, the vector field coupled 
to $\phi$ modifies the propagation speed
$c_{r2,{\rm even}}$ of $\delta \phi$. 
Along the angular direction, 
the squared propagation speed 
of $V$ in the large $l$ limit is given by Eq.~(\ref{cO3}), which is 
different from 1 in the presence of nonlinear functions of $F$ in $G_2$. 
The angular squared propagation speeds $c_{\Omega\pm,{\rm even}}^2$ 
of $\psi$ and $\delta \phi$ are expressed as Eq.~(\ref{cosqeven}) with 
$B_1$ and $B_2$ given by Eqs.~(\ref{B1def}) and (\ref{B2def}). 
These expressions of $B_1$ and 
$B_2$ are of the same forms as   those derived 
in Ref.~\cite{Kase:2021mix} without a perfect fluid, but there are 
modifications to $c_{\Omega\pm,{\rm even}}^2$ arising from the 
vector field through the term $a_4'$ [see Eq.~(\ref{conda4})]. 
We also studied the dynamics of monopole ($l=0$) and dipole ($l=1$) 
perturbations and showed that there are no additional conditions 
to those derived for $l \geq 2$.
In \hyperref[table]{Table}, we summarized all the linear stability 
conditions of odd- and even-parity perturbations. 

In Sec.~\ref{appsec}, we applied the linear stability conditions to hairy BH solutions present in Maxwell-Horndeski theories. 
In Einstein-Maxwell-dilaton theory, which is given by the action (\ref{GM}), 
there is the exact solution (\ref{GMso}) where the 
dilaton acquires a secondary hair through a coupling with the vector field. 
In this case, we showed that all the linear stability conditions are 
satisfied with luminal propagation speeds of odd- and even-parity 
perturbations. In Einstein-BI-dilaton gravity with the
action (\ref{BIa}), the angular propagation speed of vector field 
perturbation is subluminal without a scalar ghost for $\eta>0$. 
The exact BH solution (\ref{BIso}) present in Einstein-BI 
theory ($\eta=0$, $\mu(\phi)=1$) 
has neither ghost nor Laplacian instabilities. 
In Einstein-Maxwell-dilaton-GB theory with the action (\ref{GM2}), 
we showed that the hairy BH solution derived under a small $\alpha$ 
expansion can be consistent with all the linear stability conditions. 
In regularized 4DEGB gravity, however, the exact BH solution 
(\ref{fh4}) is prone to the strong coupling and instability problems.
As shown in Ref.~\cite{Tsujikawa:2022lww}, this conclusion also holds 
for an uncharged exact BH solution present in the same theory 
without the Maxwell field. 

We thus showed that the linear stability conditions derived 
in this paper are useful to exclude some BH solutions or to put 
constraints on stable parameter spaces. 
It will be of interest to apply our general framework of BH 
perturbations to the computation of quasinormal modes of BHs. 
The analysis can be also extended to the case in which 
a perfect fluid is present in Maxwell-Horndeski theories. 
This will allow us to study the stability of hairy neutron stars
along the line of Refs.~\cite{Kase:2020qvz,Kase:2020yjf,Kase:2021mix,
Minamitsuji:2022tze}. 
The scalar field coupling with 
the $U(1)$ gauge-invariant vector field can be further 
generalized by preserving the second-order property of 
equations of motion. 
Such theories are known 
as $U(1)$ gauge-invariant scalar-vector-tensor 
theories~\cite{Heisenberg:2018acv}, in which vector and scalar fields have nonminimal 
and derivative couplings to gravity. Hairy BH solutions in these theories were derived in 
Refs.~\cite{Heisenberg:2018vti,Ikeda:2019okp} and the odd-parity stability was studied 
in Refs.~\cite{Heisenberg:2018mgr,Baez:2022rdz}.
It would be also of interest to address the stability of even-parity BH perturbations in such theories. 
These issues are left for future works.

\section*{Acknowledgements}

RK is supported by the Grant-in-Aid for Young Scientists 
of the JSPS No.~20K14471. 
ST is supported by the Grant-in-Aid for Scientific Research 
Fund of the JSPS Nos.~19K03854 and 22K03642.

\appendix

\section{Coefficients in the second-order action 
of even-parity perturbations}
\label{AppA}

The coefficients in Eqs.~(\ref{Lu}) and (\ref{LdA}) are given by 
\ba
a_1&=&\sqrt{fh} \left[  \left\{ G_{4,\phi}+\frac12 h ( G_{3,X}-2 G_{4,\phi X} ) \phi'^2 \right\} r^2
+2 h \phi' \left\{ G_{4,X}-G_{5,\phi}-\frac12h ( 2 G_{4,XX}-G_{5,\phi X} ) \phi'^2 \right\} r
\right.
\notag\\
&&
\left.
+\frac12 G_{5,XX} h^3\phi'^4-\frac12 G_{5,X} h ( 3 h-1 ) \phi'^2 \right]\,, \notag\\
a_2&=&\sqrt{fh}\left( {\frac {a_1}{\sqrt{fh}}} \right)' 
- \left( {\frac {\phi''}{\phi'}}-\frac12 {\frac {f'}{f}} \right) a_1
+{\frac {r}{\phi'} \left( {\frac {f'}{f}}-{\frac {h'}{h}} \right) a_4}
+\frac{A_0'}{2}v_4\,,\qquad
a_3=-\frac12 \phi' a_1-ra_4\,,
\notag\\
a_4&=&\frac{\sqrt{fh}}{2} {\cal H}\,, \qquad
a_5=a_2'-a_1''-\left(\frac{A_0'}{2}v_4\right)'+\frac{A_0'}{2}v_5\,,\qquad
a_6=- {\frac {\sqrt {f}}{2\sqrt {h}\phi'} 
\left( {\cal H}' + \frac{{\cal H}}{r}-{\frac {{\cal F}}{r}} \right) }\,,
\notag\\
a_7 &=&a_3'-\frac{A_0'^2}{2}v_1-\frac{\phi'A_0'}{4}v_4\,,\qquad
a_8=-\frac12 {\frac {a_4}{h}}\,,\qquad
a_9= a_4'+ \left( \frac1r-\frac12 {\frac {f'}{f}} \right) a_4\,,\notag\\
b_1&=& {\frac {1}{2f}}a_4\,,\qquad
b_2=-{\frac {2}{f}}a_1\,,\qquad
b_3=- {\frac {2}{f}}(a_2-a_1')+\frac{A_0'}{f}v_4\,,\qquad
b_4=-{\frac {2}{f}}a_3\,,\qquad
b_5=-2 b_1\,,\notag\\
c_1&=&-{\frac {1}{fh}}a_1\,,\notag\\
c_2&=&\sqrt{fh} \left[  \left\{  
\frac{1}{2f}\left( -\frac12 h ( 3 G_{3,X}-8 G_{4,\phi X} ) \phi'^2
+\frac12 h^2 ( G_{3,XX}-2 G_{4,\phi XX} ) \phi'^4
-G_{4,\phi} \right) r^2
\right.\right.
\notag\\
&&
\left.\left.
-{\frac {h\phi'}{f}} \left( 
\frac12 {h^2 ( 2 G_{4,XXX}-G_{5,\phi XX} ) \phi'^4}
-\frac12 {h ( 12 G_{4,XX}-7 G_{5,\phi X} ) \phi'^2}
+3 ( G_{4,X}-G_{5,\phi} ) \right) r
\right.\right.
\notag\\
&&
\left.\left.
+\frac{h\phi'^2}{4f}\left(
G_{5,XXX} h^3\phi'^4
- G_{5,XX} h ( 10 h-1 ) \phi'^2
+3 G_{5,X}  ( 5 h-1 ) 
\right) \right\} f'
\right.
\notag\\
&&
\left.
+\phi' \left\{ \frac12G_{2,X}-G_{3,\phi}
-\frac12 h ( G_{2,XX}-G_{3,\phi X} ) \phi'^2 
+\frac{hA_0'^2}{2f}G_{2,XF}\right\} r^2
\right.
\notag\\
&&
\left.
+ 2\left\{ -\frac12h ( 3 G_{3,X}-8 G_{4,\phi X} ) \phi'^2
+\frac12h^2 ( G_{3,XX}-2 G_{4,\phi XX} ) \phi'^4
-G_{4,\phi} \right\} r
\right.
\notag\\
&&
\left.
-\frac12 h^3 ( 2 G_{4,XXX}-G_{5,\phi XX} ) \phi'^5
+\frac12 h \left\{ 2\left(6 h-1\right) G_{4,XX}+\left(1-7 h\right)G_{5,\phi X} \right\} \phi'^3
- ( 3 h-1 )  ( G_{4,X}-G_{5,\phi} ) \phi' \right] \,,\notag\\
c_3&=&-\frac12 {\frac {\sqrt {f} r^2}{\sqrt {h}}}\frac{\partial{\cal E}_{11}}{\partial\phi}\,,\notag\\
c_4&=&\frac14 \frac {\sqrt {f}}{\sqrt {h}} 
\left[ {\frac {h\phi'}{f} \left\{ 
2 G_{4,X}-2 G_{5,\phi}
-h ( 2 G_{4,XX}-G_{5,\phi X} ) \phi'^2
-{\frac {h\phi'  ( 3 G_{5,X}-G_{5,XX} \phi'^2h ) }{r}} \right\}}f'
\right.
\notag\\
&&
\left.
+4 G_{4,\phi}
+2 h ( G_{3,X}-2 G_{4,\phi X} ) \phi'^2
+{\frac {4 h ( G_{4,X}-G_{5,\phi} ) \phi'-2 h^2 ( 2 G_{4,XX}-G_{5,\phi X} ) \phi'^3}{r}} \right] \,,\notag\\
c_5&=&-h \phi'c_4-\frac12 {\frac {\sqrt{fh}}{r}}{\cal G}-\frac12 {\frac {f'}{f}}a_4\,,\notag\\
c_6&=&\frac18 {\frac {f' \phi' }{f}}a_1+\frac12 {\frac {f' r}{f}}a_4-\frac14 \phi' c_2+\frac12 h\phi' rc_4
+\frac14 \sqrt{fh}\,{\cal G}
+\frac{A_0'^2}{4}v_1+\frac{\phi'A_0'}{8}v_4\,,\notag\\
d_1&=&{\frac {1}{2f}}a_4\,,\qquad
d_2=2 hc_4\,,\notag\\
d_3&=&
-{\frac {1}{r^2} \left( {\frac {2\phi''}{\phi'}}+{\frac {h'}{h}} \right) }a_1
+{\frac {2f}{ ( f' r-2 f ) \phi'} \left( 
{\frac {2\phi''}{h\phi' r}}
+ {\frac {{f'}^{2}}{f^2}}
- {\frac {f' h'}{fh}}
-{\frac {2f'}{fr}}
+{\frac {2h'}{hr}}
+ {\frac {h'}{h^2r}} \right) }a_4
\notag\\
&&
+{\frac {f' r-2 f}{fr}}\frac{\partial a_4}{\partial\phi}
+{\frac {\sqrt {f}}{\phi' \sqrt {h}r^2}}{\cal F}
-{\frac {{f}^{3/2}}{\sqrt {h} ( f' r-2 f ) \phi'} 
\left( {\frac {f'}{fr}}+{\frac {2\phi''}{\phi' r}}+{\frac {h'}{hr}}-\frac{2}{r^2} \right) }{\cal G}
\,,\notag\\
d_4&=&\frac12 {\frac {\sqrt{fh}}{r^2}}{\cal G}
\,,\notag\\
e_1&=&{\frac {1}{\phi' fh} \left[  \left( {\frac {f'}{f}}+\frac12 {\frac {h'}{h}} \right) a_1
-2 a_1'+a_2-2 rha_6-\frac{A_0'}{2}v_4 \right] }\,,\notag\\
e_2&=&-\frac{1}{2\phi'} \left(\frac{f'}{f}a_1+2 c_2+4 hrc_4+A_0'v_4\right)\,,\qquad
e_3=\frac14 {\frac {\sqrt {f}r^2}{\sqrt {h}}}\frac{\partial{\cal E}_{\phi}}{\partial\phi}\,,\notag\\
e_4&=&{\frac {1}{\phi'}}c_4'-\frac12 {\frac {f' }{f\phi'^2h}}a_4'
-\frac12 {\frac {\sqrt {f}}{\phi'^2\sqrt {h}r}}{\cal G}'
+{\frac {1}{h\phi' r^2} \left( {\frac {\phi''}{\phi'}}+\frac12 {\frac {h'}{h}} \right) }a_1
\notag\\
&&
+{\frac {1}{4h\phi'^2} \left[ {\frac { ( f' r-6 f ) f'}{f^2r}}
+\frac {h'  ( f' r+4 f ) }{hrf}
-{\frac {4f ( 2 \phi'' h+h' \phi' ) }{\phi' h^2r ( f' r-2 f ) }} \right] }a_4
+\frac12 {\frac {h'}{h\phi'}}c_4
-\frac12 {\frac {f' r-2 f}{fhr\phi'}}\frac{\partial a_4}{\partial \phi}
\notag\\
&&
+\frac12 {\frac {f' hr-f}{r^2\sqrt {f}\phi'^2{h}^{3/2}}}{\cal F}
+\frac12 {\frac {\sqrt {f}}{r\phi'^2{h}^{3/2}} 
\left[ {\frac {f ( 2 \phi'' h+h' \phi' ) }{h\phi'  ( f' r-2 f ) }}+\frac12 {\frac {2 f-f' hr}{fr}} \right] }{\cal G}
\,,\notag\\
v_1&=& \frac{r^2}{2}\sqrt{\frac{h}{f}}\left(G_{2,F}+\frac{hA_0'^2}{f}G_{2,FF}\right)\,,\qquad
v_2=A_0'v_1\,,\qquad
v_3=-A_0'v_1-\frac{\phi'}{2}v_4\,,\qquad
v_4=-\frac{r^2h^{3/2}\phi'A_0'G_{2,XF}}{\sqrt{f}}\,,
\notag\\
v_5&=&\frac{r^2\sqrt{h}A_0'G_{2,\phi F}}{\sqrt{f}}\,,\qquad
v_6=\frac{A_0'^2}{4}v_1\,,\qquad
v_7=-2hA_0'v_8\,,\qquad
v_8=\frac{G_{2,F}}{2\sqrt{fh}}\,,\qquad
v_9=-fhv_8\,,
\ea
where ${\cal E}_{11}$ and ${\cal E}_{\phi}$ are 
defined in Eqs.~(\ref{back2}) and (\ref{Ephid}), respectively.

\section{Derivation of the GM-GHS BH solution}
\label{AppB}

In theories given by the action (\ref{GM}), GHS \cite{Garfinkle:1990qj} 
derived a static and spherically symmetric BH solution 
given by the line element 
\be
{\rm d}s^{2} =-f(\hat{r}) {\rm d}t^{2} +h^{-1}(\hat{r}) {\rm d}\hat{r}^{2} + 
\zeta^2(\hat{r}) \rd \Omega^2\,,
\label{metric2}
\ee
where $\zeta(\hat{r})$ is a function of $\hat{r}$. 
Varying the action (\ref{GM}) with respect to $f$ and $h$, 
we obtain the following two equations
\ba
& &
\zeta \zeta' h f'=f+fh \left( \phi'^2 \zeta^2-\zeta'^2 \right)
-e^{-2\phi}h A_0'^2 \zeta^2\,,
\label{GMeq1}\\
& &
\frac{f'}{f}-\frac{h'}{h}=\frac{2(\zeta''+\phi'^2 \zeta)}{\zeta'}\,,
\label{fhre}
\ea
where a prime in this Appendix \ref{AppB} represents 
the derivative with respect to $\hat{r}$.
We search for a BH solution satisfying the relation 
\be
f=h\,,
\label{fheq}
\ee
under which Eq.~(\ref{fhre}) gives 
\be
\zeta''+\phi'^2 \zeta=0\,.
\ee
Exploiting this relation for the equation of motion of $A_0'$, 
we obtain the integrated solution 
\be
A_0'=\frac{qe^{2\phi}}{\zeta^2}\,,
\label{A0dGB}
\ee
where $q$ is a constant corresponding to an electric charge.
Varying the action (\ref{GM}) with respect to $\zeta$ and $\phi$
and using the equations derived above, it follows that 
\ba
& &
\left( f \zeta^2 \right)''=2\,,\\
& &
\left( 2f \zeta^2 \phi'-f' \zeta^2 \right)'=0\,.
\ea
These equations are integrated to give 
\ba
f \zeta^2 &=& (\hat{r}-2M)^2+{\cal C}_1 (\hat{r}-2M)\,,\label{fze}\\
\phi' &=& \frac{f'}{2f}+\frac{{\cal C}_2}{2f\zeta^2}\,,
\label{phid}
\ea
where $M$, ${\cal C}_1$, and ${\cal C}_2$ are constants. 
The BH event horizon corresponds to $\hat{r}=2M$, at which 
$f=h=0$. Taking the $\hat{r}$ derivative of Eq.~(\ref{fze}) and 
using Eq.~(\ref{phid}), we obtain 
\be 
f \zeta^2 \phi'=\hat{r}-2M-f \zeta \zeta'+\frac{1}{2}({\cal C}_1
+{\cal C}_2)\,.
\label{fzephi}
\ee
Provided that $\phi'$ is finite on the horizon, the consistency of 
Eq.~(\ref{fzephi}) at $\hat{r}=2M$ requires that 
\be
{\cal C}_2=-{\cal C}_1\,.
\label{C12}
\ee
Now, we can eliminate $A_0'$, $\zeta$, $\zeta'$, and $\phi'$ in 
Eq.~(\ref{GMeq1}) by using Eqs.~(\ref{A0dGB}), (\ref{fze}), 
and (\ref{phid}) with Eq.~(\ref{C12}).
Then, we find that the metric components 
\be
f=h=1-\frac{2M}{\hat{r}}\,,\qquad 
\zeta^2=\hat{r} \left( \hat{r}-2r_q \right)
\ee
are the solutions to the above system for 
\be
{\cal C}_1=2M-2r_q\,,
\ee
where $r_q$ is defined by Eq.~(\ref{rq}). 
Integrating Eq.~(\ref{phid}) with respect to $\hat{r}$ and substituting 
the integrated solution into Eq.~(\ref{A0dGB}), the solutions to 
the scalar and vector fields are given by 
\be
\phi=\phi_0+\frac{1}{2} \ln \left( 1-\frac{2r_q}{\hat{r}} 
\right)\,,\qquad \frac{\rd A_0}{\rd \hat{r}}=\frac{qe^{2\phi_0}}{\hat{r}^2}\,,
\ee
where $\phi_0$ is the value of $\phi$ at spatial infinity. 
In the case of a magnetic charge $q$, we just need to change the 
sign of the scalar field, i.e., $\phi \to -\phi$ 
and $\phi_0 \to -\phi_0$ \cite{Garfinkle:1990qj,Gregory:1992kr,Horne:1992bi}.

To express this GM-GHS BH solution with respect to the 
line element (\ref{metric}), we perform 
the transformation 
\be
\zeta^2=\hat{r}(\hat{r}-2r_q) \to r^2\,. 
\ee
We choose the branch $\hat{r}=\sqrt{r^2+r_q^2}+r_q$
to have the property $\hat{r} \to +\infty$ as $r \to +\infty$.
Then, the above BH solution is expressed in the form (\ref{GMso}) 
for the coordinate (\ref{metric}).

\bibliographystyle{mybibstyle}
\bibliography{bib}

\end{document}